\documentclass[aps,prd,twocolumn,superscriptaddress,nofootinbib,floatfix]{revtex4-2}

\usepackage{amsmath,amssymb,bm}
\usepackage{booktabs}
\usepackage{array}
\usepackage{graphicx}
\usepackage{tikz}
\usepackage{pgfplots}
\pgfplotsset{compat=1.16}
\usetikzlibrary{patterns}
\usepackage{hyperref}
\hypersetup{colorlinks=true,citecolor=blue,linkcolor=blue,urlcolor=blue}
\usepackage{xcolor}
\usepackage{textcase}

\newcommand{\met}{\ensuremath{E_\mathrm{T}^{\rm miss}}}
\newcommand{\gev}{\ensuremath{\mathrm{GeV}}}
\newcommand{\tev}{\ensuremath{\mathrm{TeV}}}
\newcommand{\kev}{\ensuremath{\mathrm{keV}}}
\newcommand{\mev}{\ensuremath{\mathrm{MeV}}}
\newcommand{\abinv}{\ensuremath{\mathrm{ab}^{-1}}}

\newcommand{\dmz}{\ensuremath{\delta m_0}}
\newcommand{\dmp}{\ensuremath{\Delta m_+}}
\newcommand{\dmin}{\ensuremath{\Delta_{\rm min}}}

\begin{document}

\title{The LHC is not enough: \\ the LZ High-Recoil Event at FCC-hh and a Muon Collider}

\author{Benedikt Maier}
\affiliation{Department of Physics, Blackett Laboratory, Imperial College London, London SW7 2AZ, United Kingdom}
\author{Michael Spannowsky}
\affiliation{Institute for Theoretical Physics, Campus S\"ud, Karlsruhe Institute of Technology (KIT), D-76128 Karlsruhe, Germany}
\date{\today}

\begin{abstract}
The high-energy nuclear-recoil candidate reported by LUX-ZEPLIN could point to dark matter beyond the reach of the LHC. We examine how future colliders could test this interpretation, starting with the thermal Higgsino and extending to a broad class of electroweak multiplets containing neutral and charged particles. We assume an approximately pure multiplet protected by an exact stabilizing symmetry, with dominant inelastic $Z$ exchange and the full dark-matter abundance set by gauge-dominated thermal freeze-out, where mixing and additional interactions affecting production or freeze-out are negligible. In this setting, heavy thermal masses, rare production and invisible neutral transitions make the Higgsino's main collider obstacles generic across the class, independently of supersymmetry. Charged lifetimes vary, but the benchmark spectra studied, including radiative doublet splittings, remain beyond the projected HL-LHC reach. We show how FCC-hh and a high-energy muon collider can overcome these obstacles. Our FCC-hh estimates show sensitivity to the doublets and several larger multiplets through disappearing charged tracks and complementary channels that do not depend on the charged lifetime. Muon-collider projections cover the fermionic doublet and triplet at $10\,\tev$ and the full fermionic family considered here at $30\,\tev$, subject to the stated spectrum and systematic assumptions. This connects a possible underground signal to a concrete program for testing thermal electroweak dark matter at future colliders.
\end{abstract}

\maketitle

\section{Introduction}

Direct-detection experiments and colliders offer complementary ways to search for dark matter (DM). An underground detector can probe the scattering of a multi-TeV particle whose production rate at a hadron collider may be too small to observe, whereas a collider can produce and study charged partners that underground detectors cannot probe directly. A candidate signal in direct detection therefore raises the question of whether, and at which collider, its particle-physics interpretation can be tested.

The recent extended-recoil analysis of the LUX-ZEPLIN (LZ) experiment makes this question concrete. With an exposure of $2.84$ tonne-years and a nuclear-recoil window extended to about $270\,\kev$, LZ observed one event consistent with a nuclear recoil at
\begin{equation}
E_R=248\pm23\,(\mathrm{stat})\pm23\,(\mathrm{sys})\,\kev ,
\label{eq:lzevent}
\end{equation}
with a global significance of $2.6\sigma$ for the background-only hypothesis~\cite{LZ2026}. The significance is insufficient to claim discovery, but the event is informative: a conventional elastic spin-independent interpretation is strongly disfavored by the absence of the accompanying lower-energy recoils. Therefore, the event singles out dark-matter scenarios that differ qualitatively from the usual low-recoil targets.

Electroweak interpretations include approximately pure multiplets and inert scalar doublets~\cite{SmirnovEW,Nomura2026,Visinelli2026,InertDoublet2026}, while singlet--doublet mixing can reduce the transition coupling and change the charged-partner spectrum~\cite{SDscalar2026,SDfermion2026,LeeYoun2026,FanHe2026}. Supersymmetric completions~\cite{Bisal2026,Chatto2026,Frolovsky2026,LianYang2026}, radiative neutrino-mass models~\cite{Bamwidhi2026}, asymmetric dark matter~\cite{NagataYanagida2026}, and extra-dimensional constructions~\cite{LeeRandall2026} provide different origins for the masses, splittings and relic abundance. These examples motivate separating consequences of the electroweak quantum numbers from assumptions about the ultraviolet completion.

Other proposals change the interaction responsible for the recoil, through additional gauge bosons~\cite{HMLee2026,DuHuangXie2026,OkadaSeto2026,KumarPraj2026,Zhu2026}, axion portals~\cite{Unwin2026,YuanALP2026,AnAxion2026}, or transition magnetic moments~\cite{HeDipole2026,Asadi2026}; their collider rates need not follow the electroweak-multiplet pattern. Exothermic transitions~\cite{deLimaExo2026,BaerBarger2026,Xing2026} and boosted populations~\cite{Alhazmi2026,Kannike2026,MahapatraPaul2026} supply recoil energy without relying on the same high-velocity endothermic tail, while absorption~\cite{LouLu2026}, invisible nuclear transitions~\cite{AghaieStrumia2026,LeeTakahashi2026}, and neutrino scattering~\cite{Chattaraj2026} lead to still different collider questions.

Endothermic inelastic scattering~\cite{TuckerSmithWeiner} provides a simple mechanism for moving the recoil spectrum to high energies. A nearly pure Higgsino is a predictive benchmark: gauge-dominated thermal freeze-out fixes its mass near $1.1\,\tev$, and an off-diagonal $Z$ coupling can produce the xenon recoil for a neutral splitting of a few hundred keV~\cite{Freese2026,DiMauro2026,FanReece2026,Yin2026,DuWang2026,Cheung2026}.

Our target class is a single approximately pure electroweak multiplet, with negligible mixing with additional multiplets and no additional light states affecting freeze-out or production. We impose an exact $\mathbb Z_2$ under which the multiplet is odd and Standard Model fields are even; the neutral ground state is stable and a scalar multiplet has no vacuum expectation value. We assume that this state accounts for the full dark-matter density through gauge-dominated thermal freeze-out and that its inelastic nuclear scattering is dominated by $Z$ exchange. We retain the leading operators needed for the neutral and charged splittings, but neglect independent Higgs-portal interactions, including $(H^\dagger H)(\chi^\dagger\chi)$ for scalars, and other interactions that would appreciably change annihilation or collider production. The splitting interactions are likewise assumed to have negligible effects on freeze-out. These are benchmark assumptions, not properties of every inelastic-dark-matter model~\cite{Bottaro2022}.

This paper addresses the collider side of that interpretation, aspects of which have been discussed for the Higgsino in Refs.~\cite{Cheung2026,Kotlarski2026,LeYaouanc2026}. We are motivated by two observations. First, Higgsino collider studies are usually formulated within supersymmetry. We ask whether the limited LHC sensitivity follows from that framework or from the electroweak quantum numbers and spectrum alone. Second, if the interpretation cannot be tested at the LHC, it is important to know which of the proposed future colliders could test it, because a persisting signal would then bear directly on the choice of the next energy-frontier machine. 

Our results can be summarized in three steps:

\emph{(i) The Higgsino benchmark is out of reach of the HL-LHC.} The neutral splitting required by LZ is smaller than $2m_e$, so the heavier neutral state decays invisibly and the displaced-lepton searches that define the LHC inelastic-DM program~\cite{IKS,CMSiDM} do not apply. The millimeter charged-state decay length limits disappearing-track sensitivity to a few hundred GeV; inclusive recoil searches and virtual effects provide complementary probes.

\emph{(ii) The same limitations from small production rates and invisible neutral-state decays extend to the specified electroweak benchmarks.} The charged lifetimes must be assessed separately (Sec.~\ref{sec:generic}). If the scattering proceeds through $Z$ exchange, the DM is the neutral member of an electroweak multiplet with nonzero hypercharge. For the eight candidates considered here the neutral transitions are invisible, production is governed by gauge couplings, and the thermal masses lie between $0.58$ and $11.5\,\tev$. This holds true also beyond supersymmetric theories. We apply the charged-state sum rule of Ref.~\cite{Bottaro2022} to the LZ regime: within the leading splitting-operator EFT specified in Sec.~\ref{sec:spectrum}, requiring a neutral ground state bounds the lightest charged splitting. This bound and its lifetime implications depend on this EFT truncation.

\emph{(iii) FCC-hh and a multi-TeV muon collider can test these candidates.} Published FCC-hh studies exist only for the Higgsino and the wino. We extend them to the whole family with a simple model of the signal yield that retains their spin, gauge couplings, mass and charged lifetime, and which we validate against the published results (Sec.~\ref{sec:fcc}). With an expected $95\%$ confidence-level (CL) exclusion as the criterion, FCC-hh probes the radiative doublets, adopting the published reach for the fermionic case, and at the central theory inputs covers the fermionic quadruplet over approximately $70$--$167\,\mev$ in the single-branch estimate with the lower normalization and the endpoint qualification of Appendix~\ref{app:model}, covers the fermionic triplet through a Drell--Yan measurement that does not depend on the lifetime, and covers the scalar triplet, the scalar quadruplet and the fermionic quintuplet over more limited charged-splitting intervals that extend below $m_\pi$. Stable charged-particle coverage at smaller splittings is not assumed. A multi-TeV muon collider is better suited to the problem (Sec.~\ref{sec:muc}): its reach extends to a large fraction of the collision energy, and its mono-photon and mono-$W$ searches do not depend on the charged lifetime at all. Published studies of the fermionic candidates cover the doublet and the triplet at $10\,\tev$, and $\sqrt s\simeq14$ and $30\,\tev$ are needed for the quadruplet and the quintuplet; the scalar candidates have also been studied, with coverage depending on the energy, search channel and charged spectrum~\cite{Bottaro2022}.

Two caveats need to be considered. A collider cannot by itself establish that a new neutral particle makes up the Galactic dark matter, and the LZ event does not uniquely point to an electroweak multiplet. Moreover, a thermal Higgsino accounting for all of the dark matter is in tension with solar-capture limits and with the absence of events in the LZ sideband at even higher energies~\cite{Rodd2026,DiMauroShaikh2026,Bose2026,PospelovRamani2026,GhoshChavezKelso2026,Langhoff2026}. We therefore use the event only as a motivation for a sharply defined class of collider targets.

The paper is organized as follows. Section~\ref{sec:kin} explains why the event points to inelastic scattering and what a single recoil can and cannot determine. Section~\ref{sec:higgsino} reviews the Higgsino benchmark, its LHC coverage and its noncollider constraints. Section~\ref{sec:generic} extends the discussion to the specified electroweak multiplets. Sections~\ref{sec:fcc} and~\ref{sec:muc} assess FCC-hh and a muon collider. Section~\ref{sec:caveats} collects caveats and an outlook, and Sec.~\ref{sec:conc} concludes. Two appendices contain the supersymmetric charged spectrum and the details of our FCC-hh yield model.

\section{Why the event points to inelastic dark matter}
\label{sec:kin}

We first derive the kinematic constraints on the dark-matter mass and neutral splitting from the observed recoil energy.

\subsection{Endothermic kinematics}

Consider a halo particle $\chi_1$ that scatters off a nucleus $N$ into a slightly heavier state,
\begin{equation}
\chi_1+N\to \chi_2+N,
\qquad \delta\equiv m_{\chi_2}-m_{\chi_1}>0 .
\end{equation}
Part of the kinetic energy is used up to create the heavier state, so a recoil of energy $E_R$ requires a minimum incident speed
\begin{equation}
 v_{\min}(E_R)=\frac{1}{\sqrt{2m_NE_R}}
 \left(\frac{m_NE_R}{\mu_{\chi N}}+\delta\right),
\label{eq:vmin}
\end{equation}
where $m_N$ is the nuclear mass and $\mu_{\chi N}$ the reduced mass. For elastic scattering ($\delta=0$) the required speed grows with $E_R$, so the spectrum is largest at the lowest recoil energies. This is why the elastic interpretation fails: a spin-independent elastic spectrum normalized to one event at $248\,\kev$ predicts of order $10^4$ events below $55\,\kev$, where LZ sees nothing beyond background~\cite{LZ2026}. For $\delta>0$ the situation is reversed. The second term in Eq.~(\ref{eq:vmin}) diverges at small $E_R$, so low-energy recoils are forbidden altogether, and scattering is possible only for particles faster than
\begin{equation}
 v_{\rm thr}=\sqrt{\frac{2\delta}{\mu_{\chi N}}},
 \qquad\text{i.e.}\qquad
 \delta\leq \tfrac12\mu_{\chi N}v_{\max}^2 ,
\label{eq:deltamax}
\end{equation}
where the second form follows from requiring $v_{\rm thr}\le v_{\max}$, with $v_{\max}$ the largest DM speed in the detector frame: if even the fastest halo particle is below threshold, no scattering occurs at all. For a standard halo with a Maxwellian velocity distribution of width $v_0=238\,\mathrm{km/s}$, truncated at the Galactic escape speed $v_{\rm esc}=544\,\mathrm{km/s}$ and boosted into the detector frame by the laboratory velocity $v_{\rm lab}=250\,\mathrm{km/s}$, this is $v_{\max}=v_{\rm esc}+v_{\rm lab}\simeq790\,\mathrm{km/s}$; we quote results for $770$--$800\,\mathrm{km/s}$ to indicate the sensitivity to this assumption. An endothermic splitting thus removes the low-recoil events and leaves a spectrum concentrated at high energies, exactly the pattern suggested by the LZ candidate. Momentum-dependent and spin-dependent elastic interactions can also shift events upward and remain marginally viable~\cite{LZ2026,ElahiSchwaller2026,Arcadi2026}; they lead to a different collider phenomenology and are not pursued here.

For xenon and $m_\chi=1.1\,\tev$ the reduced mass is $\mu_{\chi N}\simeq110\,\gev$, and Eq.~(\ref{eq:deltamax}) gives a largest possible splitting of $360$--$390\,\kev$ for $v_{\max}=770$--$800\,\mathrm{km/s}$. Splittings of a few hundred keV are therefore right at the limit of what the fastest halo particles can excite. A larger splitting is not impossible, but it requires a faster population than the standard halo provides.

\subsection{Constraints from the LZ candidate event}

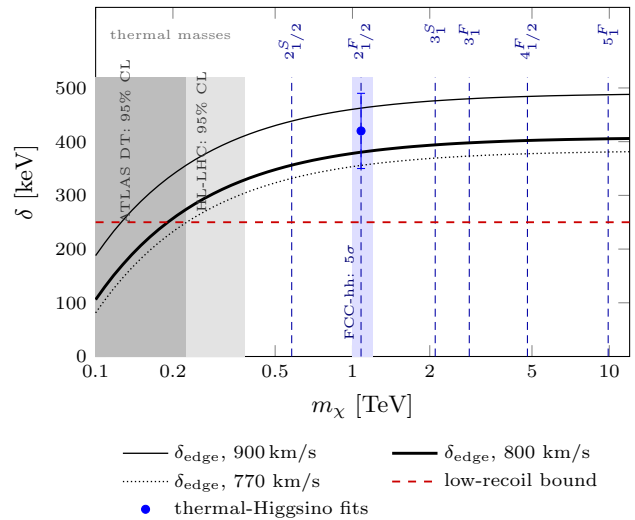
\begin{figure}[t]
\centering
\begin{tikzpicture}
\begin{axis}[
  width=1.0\columnwidth, height=6.2cm,
  xmode=log, xmin=0.1, xmax=12, ymin=0, ymax=650,
  xlabel={$m_\chi$ [TeV]}, ylabel={$\delta$ [keV]},
  xtick={0.1,0.2,0.5,1,2,5,10}, xticklabels={0.1,0.2,0.5,1,2,5,10},
  ytick={0,100,200,300,400,500},
  log ticks with fixed point,
  tick label style={font=\scriptsize}, label style={font=\small},
  legend style={at={(0.5,-0.22)},anchor=north,font=\scriptsize,draw=none,fill=none,
     /tikz/every even column/.append style={column sep=0.22cm}},
  legend columns=2, legend cell align=left, clip=true]
\fill[black!26] (axis cs:0.1,0) rectangle (axis cs:0.225,520);
\fill[black!11] (axis cs:0.225,0) rectangle (axis cs:0.38,520);
\fill[blue!13] (axis cs:1.0,0) rectangle (axis cs:1.2,520);
\draw[densely dashed,blue!65!black,line width=0.4pt] (axis cs:0.58,0) -- (axis cs:0.58,520);
\draw[densely dashed,blue!65!black,line width=0.4pt] (axis cs:1.08,0) -- (axis cs:1.08,520);
\draw[densely dashed,blue!65!black,line width=0.4pt] (axis cs:2.1,0) -- (axis cs:2.1,520);
\draw[densely dashed,blue!65!black,line width=0.4pt] (axis cs:2.85,0) -- (axis cs:2.85,520);
\draw[densely dashed,blue!65!black,line width=0.4pt] (axis cs:4.8,0) -- (axis cs:4.8,520);
\draw[densely dashed,blue!65!black,line width=0.4pt] (axis cs:9.9,0) -- (axis cs:9.9,520);
\node[font=\tiny,rotate=90,anchor=east,blue!55!black] at (axis cs:0.58,640) {$2^{S}_{1/2}$};
\node[font=\tiny,rotate=90,anchor=east,blue!55!black] at (axis cs:1.08,640) {$2^{F}_{1/2}$};
\node[font=\tiny,rotate=90,anchor=east,blue!55!black] at (axis cs:2.1,640)  {$3^{S}_{1}$};
\node[font=\tiny,rotate=90,anchor=east,blue!55!black] at (axis cs:2.85,640) {$3^{F}_{1}$};
\node[font=\tiny,rotate=90,anchor=east,blue!55!black] at (axis cs:4.8,640)  {$4^{F}_{1/2}$};
\node[font=\tiny,rotate=90,anchor=east,blue!55!black] at (axis cs:9.9,640)  {$5^{F}_{1}$};
\node[font=\tiny,anchor=west,black!55] at (axis cs:0.105,600) {thermal masses};
\addplot[black,line width=0.5pt] coordinates {(0.1,187.9) (0.1085,211.5) (0.1176,233.3) (0.1276,253.3) (0.1383,271.8) (0.15,288.8) (0.1627,304.5) (0.1765,319.0) (0.1914,332.4) (0.2076,344.7) (0.2251,356.1) (0.2441,366.6) (0.2648,376.2) (0.2872,385.1) (0.3114,393.3) (0.3378,400.9) (0.3663,407.9) (0.3973,414.3) (0.4309,420.3) (0.4673,425.7) (0.5068,430.8) (0.5496,435.4) (0.5961,439.7) (0.6464,443.7) (0.7011,447.3) (0.7603,450.7) (0.8246,453.8) (0.8943,456.7) (0.9699,459.3) (1.052,461.7) (1.141,464.0) (1.237,466.0) (1.342,467.9) (1.455,469.7) (1.578,471.3) (1.712,472.8) (1.856,474.2) (2.013,475.5) (2.183,476.6) (2.368,477.7) (2.568,478.7) (2.785,479.6) (3.021,480.5) (3.276,481.3) (3.553,482.0) (3.853,482.6) (4.179,483.2) (4.532,483.8) (4.915,484.3) (5.331,484.8) (5.781,485.3) (6.27,485.7) (6.8,486.0) (7.375,486.4) (7.998,486.7) (8.674,487.0) (9.407,487.3) (10.2,487.5) (11.06,487.8) (12,488.0)};
\addlegendentry{$\delta_{\rm edge}$, $900$\,km/s}
\addplot[black,line width=1.1pt] coordinates {(0.1,105.9) (0.1085,129.5) (0.1176,151.2) (0.1276,171.2) (0.1383,189.7) (0.15,206.8) (0.1627,222.5) (0.1765,237.0) (0.1914,250.3) (0.2076,262.7) (0.2251,274.0) (0.2441,284.5) (0.2648,294.2) (0.2872,303.1) (0.3114,311.3) (0.3378,318.9) (0.3663,325.8) (0.3973,332.3) (0.4309,338.2) (0.4673,343.7) (0.5068,348.7) (0.5496,353.4) (0.5961,357.7) (0.6464,361.6) (0.7011,365.3) (0.7603,368.6) (0.8246,371.7) (0.8943,374.6) (0.9699,377.2) (1.052,379.7) (1.141,381.9) (1.237,384.0) (1.342,385.9) (1.455,387.6) (1.578,389.3) (1.712,390.8) (1.856,392.1) (2.013,393.4) (2.183,394.6) (2.368,395.7) (2.568,396.7) (2.785,397.6) (3.021,398.4) (3.276,399.2) (3.553,399.9) (3.853,400.6) (4.179,401.2) (4.532,401.8) (4.915,402.3) (5.331,402.8) (5.781,403.2) (6.27,403.6) (6.8,404.0) (7.375,404.3) (7.998,404.6) (8.674,404.9) (9.407,405.2) (10.2,405.5) (11.06,405.7) (12,405.9)};
\addlegendentry{$\delta_{\rm edge}$, $800$ km/s}
\addplot[black,line width=0.5pt,densely dotted] coordinates {(0.1,81.3) (0.1085,104.8) (0.1176,126.6) (0.1276,146.6) (0.1383,165.1) (0.15,182.2) (0.1627,197.9) (0.1765,212.4) (0.1914,225.7) (0.2076,238.1) (0.2251,249.4) (0.2441,259.9) (0.2648,269.5) (0.2872,278.5) (0.3114,286.7) (0.3378,294.2) (0.3663,301.2) (0.3973,307.7) (0.4309,313.6) (0.4673,319.1) (0.5068,324.1) (0.5496,328.8) (0.5961,333.1) (0.6464,337.0) (0.7011,340.7) (0.7603,344.0) (0.8246,347.1) (0.8943,350.0) (0.9699,352.6) (1.052,355.1) (1.141,357.3) (1.237,359.4) (1.342,361.3) (1.455,363.0) (1.578,364.6) (1.712,366.1) (1.856,367.5) (2.013,368.8) (2.183,370.0) (2.368,371.0) (2.568,372.0) (2.785,373.0) (3.021,373.8) (3.276,374.6) (3.553,375.3) (3.853,376.0) (4.179,376.6) (4.532,377.1) (4.915,377.7) (5.331,378.1) (5.781,378.6) (6.27,379.0) (6.8,379.4) (7.375,379.7) (7.998,380.0) (8.674,380.3) (9.407,380.6) (10.2,380.9) (11.06,381.1) (12,381.3)};
\addlegendentry{$\delta_{\rm edge}$, $770$ km/s}
\addplot[red!80!black,dashed,line width=0.7pt] coordinates {(0.1,250) (12,250)};
\addlegendentry{low-recoil bound}
\addplot[only marks,mark=*,mark size=1.5pt,blue,error bars/.cd,y dir=both,y explicit]
  coordinates {(1.08,420) +- (0,70)};
\addlegendentry{thermal-Higgsino fits}
\node[font=\tiny,rotate=90,anchor=south,black!80] at (axis cs:0.148,400) {ATLAS DT: $95\%$ CL};
\node[font=\tiny,rotate=90,anchor=south,black!80] at (axis cs:0.293,400) {HL-LHC: $95\%$ CL};
\node[font=\tiny,rotate=90,anchor=south,blue!45!black] at (axis cs:1.095,120) {FCC-hh: $5\sigma$};
\end{axis}
\end{tikzpicture}
\caption{What a single recoil at $E_R=248\,\kev$ on xenon can tell us. The curves show the largest splitting $\delta_{\rm edge}(m_\chi)$, Eq.~(\ref{eq:edge}), for which the event can be produced at all, for three values of the maximal DM speed in the detector frame; the event is kinematically possible only \emph{below} the relevant curve. The curves are not likelihood contours and contain no information from the sideband, solar or halo analyses discussed in Sec.~\ref{sec:noncollider}. Vertical dashed lines mark the thermal masses of the electroweak candidates of Sec.~\ref{sec:generic}, labelled $n^{F,S}_{Y}$ for fermions and scalars~\cite{Bottaro2022}. The dashed red line is the approximate lower bound on the splitting of a $Z$-coupled state from low-recoil data~\cite{NagataShirai,Bottaro2022}, and the point with error bar spans the published thermal-Higgsino fits~\cite{DiMauro2026,FanReece2026,Rodd2026}. The shading refers to the radiative fermionic-doublet benchmark: ATLAS excludes masses below $225\,\gev$ at $95\%$ CL~\cite{ATLAS2026}, and the lighter gray extension indicates projected HL-LHC exclusions up to $250$--$380\,\gev$~\cite{HLLHC,MSZ}. The blue band instead marks published $5\sigma$ discovery reaches of $1.0$--$1.2\,\tev$ across the $80$--$100\,\tev$ FCC-hh energy and detector scenarios~\cite{Saito2019,FCCFSR2025,FCCPED2025}; it is not a $95\%$ exclusion band. These collider scales do not constrain the neutral splitting independently.}
\label{fig:plane}
\end{figure}

Requiring that the observed recoil can be produced at all, $v_{\min}(E_R)\le v_{\max}$, gives the largest splitting compatible with the event at a given mass,
\begin{equation}
 \delta_{\rm edge}(m_\chi)=v_{\max}\sqrt{2m_NE_R}-E_R-\frac{m_NE_R}{m_\chi},
\label{eq:edge}
\end{equation}
where $\sqrt{2m_NE_R}\simeq246\,\mev$ is the momentum transfer of the event. Figure~\ref{fig:plane} shows this kinematic edge for three values of $v_{\max}$. This has two implications.

First, the LZ event constrains the mass only loosely. For $m_\chi\gg m_N$ the last term in Eq.~(\ref{eq:edge}) is only $28\,\kev$ at $1.1\,\tev$ and $3\,\kev$ at $10\,\tev$, so a TeV particle and a ten-TeV particle are kinematically indistinguishable. The LZ likelihood itself only requires $m_\chi\gtrsim100\,\gev$~\cite{LZ2026}. The mass scale of the collider target must therefore come from elsewhere; in this paper it comes from the requirement that the particle is a thermal relic (Sec.~\ref{sec:generic}).

Second, for the splittings of interest the event must come from the extreme tail of the velocity distribution. At $m_\chi=1.1\,\tev$ a splitting of $350\,\kev$ requires $v_{\min}\simeq760\,\mathrm{km/s}$, whereas elastic scattering at the same mass requires $340\,\mathrm{km/s}$. For the standard halo defined above, only a fraction $10^{-3}$ of the particles is faster than $750\,\mathrm{km/s}$ and only $6\times10^{-6}$ is faster than $790\,\mathrm{km/s}$. This strong suppression is what allows an interaction of full electroweak strength to produce a single event in several tonne-years~\cite{FanReece2026}, and it is the reason why the inferred splitting is unusually sensitive to the local escape speed, to halo substructure and to the Large Magellanic Cloud~\cite{FanReece2026,GhoshChavezKelso2026,McCabe2026,Chujo2026}.

\section{The Higgsino benchmark}
\label{sec:higgsino}

We first review the interpretation that dominates the recent literature. Our aim is to identify which of its features matter for colliders, whose generality we then assess in Sec.~\ref{sec:generic}.

\subsection{Why a Higgsino fits the event}

In supersymmetry the Higgsinos are the fermionic partners of the two Higgs doublets; they form the $(2,\tfrac12)$ electroweak multiplet in Sec.~\ref{sec:generic}. Their common mass is the parameter $\mu$, while the partners of the gauge bosons, the bino and the wino, have masses $M_1$ and $M_2$. If the latter are heavy, $|M_1|,|M_2|\gg|\mu|$, the light states are two neutral Higgsinos $\widetilde\chi^0_{1,2}$ and a charged one, $\widetilde\chi^\pm_1$. The two neutral states would form a single Dirac fermion if dark-matter number were conserved; a small violation of it splits them into two Majorana states. The coupling to the $Z$ boson then connects the two \emph{different} mass eigenstates,
\begin{equation}
 \mathcal L_Z\supset \frac{g}{2c_W}\,Z_\mu\,
 i\,\overline{\widetilde\chi_1^0}\gamma^\mu\widetilde\chi_2^0 ,
\label{eq:Zcoupling}
\end{equation}
because the vector current of a single Majorana fermion vanishes identically. Scattering through $Z$ exchange is therefore automatically inelastic.

This structure explains why a splitting is not optional. For the neutral member of a multiplet with hypercharge $Y$, unsuppressed $Z$ exchange would give a spin-independent cross section on neutrons of
\begin{equation}
 \sigma_n^{Z}=\frac{2G_F^2\mu_n^2Y^2}{\pi}
 \simeq 7\times10^{-39}\,\mathrm{cm}^2\times(2Y)^2 ,
\label{eq:sigmaZ}
\end{equation}
nearly eight orders of magnitude above the LZ limit at a TeV~\cite{LZ2024}. A Higgsino survives direct detection only because the splitting makes the scattering kinematically impossible for almost all halo particles~\cite{NagataShirai}. The same observation turns the LZ event into a measurement. At fixed mass, local density and halo distribution, the gauge coupling in Eq.~(\ref{eq:Zcoupling}) leaves the neutral splitting as the parameter suppressing the rate. An expected signal yield of order one event then places the splitting close to the kinematic edge of Fig.~\ref{fig:plane}, where only the extreme velocity tail contributes. The preferred splitting is therefore conditional on these inputs and the recoil likelihood,
\begin{equation}
 \dmz\equiv m_{\widetilde\chi_2^0}-m_{\widetilde\chi_1^0}\simeq0.35\text{--}0.5~\mev ,
\label{eq:neutral_split}
\end{equation}
where $\dmz$ is the same splitting called $\delta$ in Sec.~\ref{sec:kin}, renamed here because a second splitting will appear shortly. The quoted range reflects different assumptions about the velocity tail~\cite{Freese2026,DiMauro2026,FanReece2026,Rodd2026}; its upper part lies above the edge of Fig.~\ref{fig:plane} for $v_{\max}=800\,\mathrm{km/s}$ and requires a faster population than the standard halo, for instance from the Large Magellanic Cloud. The mass is fixed independently by thermal freeze-out, $m_{\widetilde H}\simeq1.0$--$1.1\,\tev$~\cite{CST,Bottaro2022}.

\subsection{Two splittings with different roles}

Two mass splittings appear in the problem. The neutral splitting $\dmz$ of Eq.~(\ref{eq:neutral_split}) controls what LZ sees. What a collider sees is controlled by the splitting between the charged and the lightest neutral state,
\begin{equation}
 \dmp\equiv m_{\widetilde\chi_1^\pm}-m_{\widetilde\chi_1^0}\simeq0.34\text{--}0.36\,\gev ,
\end{equation}
which is generated by electroweak loops and is a thousand times larger~\cite{ThomasWells,CFS}. It determines the lifetime of the charged state, which decays predominantly through $\widetilde\chi_1^\pm\to\widetilde\chi^0\pi^\pm$ with the width
\begin{equation}
 \Gamma_\pi\simeq
 \frac{G_F^2|V_{ud}|^2f_\pi^2}{\pi}
 (\dmp)^3
 \sqrt{1-\frac{m_\pi^2}{(\dmp)^2}},
\label{eq:pionwidth}
\end{equation}
where $f_\pi\simeq130\,\mev$. Including the leptonic modes, which add about $9\%$ to the pion width in this range (Appendix~\ref{app:model}), one finds
\begin{equation}
 c\tau_{\widetilde\chi_1^\pm}\simeq 6\text{--}7\,\mathrm{mm},
\label{eq:lifetime}
\end{equation}
in agreement with the values used in the experimental studies~\cite{Saito2019,ATLAS2026}. Table~\ref{tab:benchmark} summarizes the benchmark. The charged state travels a few millimeters and then decays into an invisible neutral particle and a pion that is too soft to be reconstructed. This defines the signature at a collider.

\begin{table}[t]
\caption{Characteristic scales of the thermal-Higgsino interpretation of the LZ event. The neutral splitting is inferred from the event rate and depends on the assumed halo; the charged splitting and the lifetime are predictions of the pure-doublet limit.}
\label{tab:benchmark}
\begin{ruledtabular}
\begin{tabular}{lc}
Quantity & Value \\
\hline
Mass $m_{\widetilde H}$ & $1.0$--$1.1\,\tev$ \\
Recoil energy $E_R$ & $248\,\kev$ \\
Neutral splitting $\dmz$ & $0.35$--$0.5\,\mev$ \\
Charged splitting $\dmp$ & $0.34$--$0.36\,\gev$ \\
Charged decay length $c\tau$ & $6$--$7\,\mathrm{mm}$ \\
\end{tabular}
\end{ruledtabular}
\end{table}

\subsection{Why the LHC cannot test it}
\label{sec:lhc}

Four properties limit the LHC sensitivity to the radiative Higgsino benchmark: (a) its TeV mass and (b) its electroweak production rate, (c) its invisible neutral transitions, and (d) the short charged-state decay length. The first three motivate the comparison with other multiplets in Sec.~\ref{sec:generic}; the fourth depends on their charged spectra.

\emph{The neutral sector is invisible.} The LHC searches for inelastic DM look for the decay $\chi_2\to\chi_1\ell^+\ell^-$, which produces soft, displaced lepton pairs~\cite{IKS,CMSiDM}. The LZ splitting lies below $2m_e=1.02\,\mev$, so this decay is kinematically closed. The heavier neutral state can only decay into neutrinos or into a photon with an energy of a few hundred keV, far below any calorimeter threshold, and for a pure multiplet it is long-lived on detector scales in any case~\cite{Bottaro2022}. Both neutral states escape detection and contribute to missing transverse momentum. Charged-state tracks, neutral-pair production accompanied by visible radiation, and virtual corrections to Standard Model processes remain possible collider probes.

\emph{The charged partner is produced rarely and decays early.} The relevant search looks for a ``disappearing track'': a charged particle that crosses a few tracker layers and then vanishes. The most recent ATLAS analysis requires at least three to four pixel hits, corresponding to a flight distance of roughly $10\,\mathrm{cm}$, and excludes pure Higgsinos only up to
\begin{equation}
 m_{\widetilde\chi_1^\pm}\simeq 225\,\gev
\label{eq:atlas}
\end{equation}
for lifetimes below $0.03\,\mathrm{ns}$, that is for $c\tau\lesssim9\,\mathrm{mm}$, which is the range that contains the Higgsino. For a lifetime near $1\,\mathrm{ns}$, corresponding to a track of some $30\,\mathrm{cm}$, the same search reaches about $720\,\gev$~\cite{ATLAS2026}. The reason is twofold. Production proceeds only through electroweak interactions and falls steeply with mass; at $1.1\,\tev$ the total cross section at $13\,\tev$ is below $1\,\mathrm{fb}$~\cite{Cheung2026}. In addition, with a decay length of a few millimeters only the rare charginos with a very large boost survive long enough to leave a track, which forces the selection into the tail of hard initial-state radiation. Projections for the HL-LHC place the $95\%$~CL reach for the pure Higgsino near $250\,\gev$~\cite{HLLHC}, or up to about $380\,\gev$ with an optimistic forward tracker~\cite{MSZ}; soft-track and monojet strategies reach a similar range~\cite{FNOS,FNOOS,HKMM,LowWang}. Soft-lepton searches relying on the neutral dilepton decay do not apply. Charged decays into $\chi^0\ell\nu$ remain allowed, but their available energy in this benchmark is only a few hundred MeV. LEP constrains such compressed states up to about $90\,\gev$~\cite{LEPSUSY}.

The thermal Higgsino at $1.1\,\tev$ is therefore a factor of three to four beyond what the HL-LHC can exclude. The limited reach follows from the combination of a TeV-scale mass, electroweak production and a millimeter decay length. Longer-lived electroweakinos are constrained at substantially higher masses.

\subsection{Noncollider constraints}
\label{sec:noncollider}

Because the interpretation will be judged first by noncollider data, we briefly summarize which constraints the inelastic splitting removes and which it does not.

The splitting removes the conventional direct-detection limits. For $\dmz\gtrsim0.25\,\mev$ the low-recoil windows of LZ, XENONnT and PandaX-4T are kinematically closed at TeV masses, so the strong spin-independent limits do not apply~\cite{NagataShirai,Bottaro2022}; the irreducible loop-induced elastic scattering is far below current sensitivity. Indirect detection is not affected by the splitting to a good approximation, and current gamma-ray data do not yet exclude the thermal Higgsino, although CTAO is expected to reach it~\cite{Dessert2023,Rinchiuso2021,WuZhangZhu2026}.

Dark matter falling into the Sun is accelerated far beyond Galactic speeds, so the inelastic threshold is easily overcome, and the subsequent annihilation produces neutrinos that Super-Kamiokande and IceCube constrain. Several analyses conclude that a full-density thermal Higgsino is compatible with these limits only for $\dmz\gtrsim0.51$--$0.56\,\mev$, above the range preferred by the LZ event~\cite{DiMauroShaikh2026,Bose2026,PospelovRamani2026,NguyenLindenHooper2026}, and that halo uncertainties are unlikely to remove the conflict~\cite{GhoshChavezKelso2026}. A nonthermal treatment of the captured population can weaken the bound~\cite{QiSun2026}. A second concern is internal to LZ: the same near-threshold kinematics that produces the event predicts a recoil spectrum that extends up to about $600\,\kev$, and the absence of events in the sideband above the analysis window has been argued to be in tension with the Higgsino interpretation~\cite{Rodd2026,DentNewstead2026,Langhoff2026}.

These results weaken a literal reading of the event as the discovery of a thermal Higgsino. They do not affect the collider question posed here, which concerns a well-defined hypothesis that may have to be confirmed or refuted; we return to them in Sec.~\ref{sec:caveats}.

\section{Electroweak multiplet benchmarks}
\label{sec:generic}

The radiative Higgsino benchmark combines electroweak production, a TeV-scale mass, invisible neutral transitions and a short charged-state lifetime. Under the assumptions stated in the Introduction, gauge-dominated production and an invisible neutral transition extend to the other multiplets, whose thermal masses range from sub-TeV to multi-TeV values. Their charged lifetimes do not follow universally from the Higgsino example: the larger multiplets can have long-lived charged states. We therefore discuss the spectrum and the relevant search channels separately.

\subsection{The class of candidates}
\label{sec:class}

For the dominant $Z$-exchange interaction assumed here, the DM must be the neutral component of an electroweak multiplet with hypercharge $Y\neq0$; multiplets with $Y=0$, such as the wino, have no diagonal neutral $Z$ coupling at tree level and are outside this target class. We therefore consider a complex $n$-plet of $SU(2)_L$ with $Y\neq0$, either a Dirac fermion or a complex scalar, whose neutral component is split into two nearly degenerate states~\cite{CFS,Bottaro2022}. With negligible mixing, the $Z$ coupling is fixed by $Y$ and Eq.~(\ref{eq:sigmaZ}) applies. Inferring a preferred neutral splitting additionally requires a halo model, a local density and the recoil likelihood; the kinematic edge alone does not establish an acceptable event-rate fit for every candidate.

To define a collider target we need the mass, which the event does not provide (Sec.~\ref{sec:kin}). We assume that the multiplet is a thermal relic whose abundance is set by its gauge interactions. The mass is then a prediction, which we call the thermal mass $m_{\rm th}$, and the complete list of candidates is known~\cite{Bottaro2021,Bottaro2022}: the multiplets $n_{1/2}$ with $n=2,4,\dots,12$, together with $3_1$ and $5_1$, for both spins. We label a multiplet by its dimension and hypercharge, $(n,Y)$, abbreviated $n_Y$ in the figures, and the fermionic doublet $(2,\tfrac12)$ is the Higgsino of Sec.~\ref{sec:higgsino}. Table~\ref{tab:multiplets} lists the eight candidates with thermal masses below $12\,\tev$. At a $100\,\tev$ proton collider, $\hat s=x_1x_2s\leq s$, so the partonic center-of-mass energy cannot exceed $100\,\tev$. The fermionic and scalar sextuplet central thermal masses are $31.8$ and $32.7\,\tev$, giving pair thresholds of $63.6$ and $65.4\,\tev$, respectively~\cite{Bottaro2022}. They are kinematically accessible in principle, although production requires very large parton momentum fractions. The $n\geq8$ thermal candidates in that classification have masses above $80\,\tev$ and cannot be pair-produced at $100\,\tev$. We leave the sextuplets outside our numerical study and do not claim pair-production access to all higher representations. The same class has been discussed under many names---the Higgsino, the vectorlike lepton doublet~\cite{ThomasWells}, Minimal Dark Matter~\cite{CFS}, complex WIMPs~\cite{Bottaro2022}, the inert doublet~\cite{ArinaLingTytgat}, and the doublet limit of singlet--doublet models~\cite{SingletDoublet}---and several of these have already been applied to the LZ event without reference to supersymmetry~\cite{SmirnovEW,AhmedLeontaris2026,SDscalar2026,SDfermion2026,InertDoublet2026,Nomura2026,Visinelli2026,LeeYoun2026,FanHe2026}.

\begin{table*}[t]
\caption{Thermal electroweak benchmarks (Dirac fermions or complex scalars) and indicative $95\%$ CL collider coverage. Masses follow Ref.~\cite{Bottaro2022}; $\dmin$ and $c\tau$ denote the lightest charged splitting and proper decay length. The kinematic edge $\delta_{\rm edge}$ uses Eq.~(\ref{eq:edge}) with $v_{\max}=800\,\mathrm{km/s}$. FCC-hh assumes $100\,\tev$ and $30\,\abinv$; track intervals (GeV) use central inputs and the lower single-branch normalization, with qualifications in Table~\ref{tab:windows} and Appendix~\ref{app:model}. Muon-collider energies denote published reaches, not production thresholds. DY: Drell--Yan; DT: disappearing tracks; MIM: missing invariant mass. $^a$Radiative splitting, adjustable by additional interactions. $^b$Quartic scalar interactions modify the mass splittings. $^c$EFT bound (\ref{eq:bound2}); charged-state lifetimes. $^d$Percent-level background systematics. $^e$Published scalar benchmarks (Sec.~\ref{sec:muc}); DT depends on the spectrum, and triplet MIM requires sub-percent systematics. $^f$Recoil reach requires sub-percent systematics; otherwise use tracks. $^g$HL-LHC benchmark spectra only, including radiative doublet splittings (Sec.~\ref{sec:generic}).}
\label{tab:multiplets}
\footnotesize
\begin{ruledtabular}
\begin{tabular}{llcccclll}
$(n,Y)$ & Spin & $m_{\rm th}$ [TeV] & $\dmin$ [GeV] & $c\tau_{\chi^\pm}$ & $\delta_{\rm edge}$ [keV] & HL-LHC$^{\,g}$ & FCC-hh & $\mu$ collider, $\sqrt s$ \\
\hline
$(2,\tfrac12)$ & scalar & $0.58\pm0.01$ & $0.35^{\,a,b}$ & $6.4$ mm & $356$ & no & tracks; DY$^{\,d}$ & $6\,\tev$, DT$^{\,e}$ \\
$(2,\tfrac12)$ & fermion & $1.08\pm0.02$ & $0.35^{\,a}$ & $6.4$\,mm & $380$ & no & tracks, DY, mono-$j^{\,d}$ & ${\ge}3\,\tev$ \\
$(3,1)$ & scalar & $2.1\pm0.1$ & $0.54^{\,a,b}$ & $0.75$\,mm & $394$ & no & tracks: $0.080$--$0.228$ & $14\,\tev$, MIM$^{\,e}$ \\
$(3,1)$ & fermion & $2.85\pm0.14$ & $0.54^{\,a}$ & $0.75$\,mm & $398$ & no & DY, mono-$j^{\,d}$; tracks: $0.056$--$0.229$ & ${\ge}10\,\tev$ \\
$(4,\tfrac12)$ & fermion & $4.8\pm0.3$ & ${\le}0.17^{\,c}$ & ${\ge}27$\,mm & $402$ & no & tracks: $0.070$--$0.167$ & ${\ge}14\,\tev^{\,f}$ \\
$(4,\tfrac12)$ & scalar & $4.98\pm0.25$ & ${\le}0.17^{\,c}$ & ${\ge}27$\,mm & $402$ & no & tracks: $0.130$--$0.148$ & $14\,\tev$, DT$^{\,e}$ \\
$(5,1)$ & fermion & $9.9\pm0.7$ & ${\le}0.17^{\,c}$ & ${\ge}18$\,mm & $405$ & no & tracks: $0.125$--$0.140$ & ${\gtrsim}20\,\tev$ \\
$(5,1)$ & scalar & $11.5\pm0.8$ & ${\le}0.17^{\,c}$ & ${\ge}18$\,mm & $406$ & no & no & not at $30\,\tev^{\,e}$ \\
\end{tabular}
\end{ruledtabular}
\end{table*}

\subsection{Mass, production and the invisible neutral sector}

Three features follow within the benchmark assumptions. \emph{(a) Mass.} The thermal masses in Table~\ref{tab:multiplets} range from $0.58\,\tev$ for the scalar doublet to $11.5\,\tev$ for the scalar quintuplet; they are fixed by gauge interactions, including Sommerfeld enhancement and bound-state formation~\cite{CST,Bottaro2021,Bottaro2022}. \emph{(b) Production.} Pair production through $W$, $Z$ and $\gamma$ is fixed by $n$ and $Y$. Additional production interactions are neglected by assumption. \emph{(c) Invisibility.} Figure~\ref{fig:plane} shows that the kinematic edge saturates near $0.4\,\mev$ for all masses, so the neutral splitting is below $2m_e$ for every candidate. As for the Higgsino, the heavier neutral state decays only into neutrinos or, for fermions, into a soft photon, and searches relying on the neutral dilepton decay are inapplicable. For a scalar the photon channel is absent and the lifetime is of order $10^6\,\mathrm{s}$; since each decay releases only the splitting energy, a fraction $\dmz/m\sim10^{-6}$ of the DM energy density and entirely in neutrinos, we regard the requirement of a decay before nucleosynthesis imposed in Ref.~\cite{Bottaro2022} as conservative, but a dedicated study would be worthwhile.

\subsection{Charged-spectrum bounds}
\label{sec:spectrum}

\begin{figure}[t]
\centering
\begin{tikzpicture}
\begin{axis}[
  width=1.0\columnwidth, height=6.4cm,
  ymode=log, xmin=0.04, xmax=0.80, ymin=0.05, ymax=1e7,
  xlabel={$\Delta_{\rm min}=m_{\chi^\pm}-m_{\chi^0}$ [GeV]},
  ylabel={$c\tau_{\chi^\pm}$ [mm]},
  ytick={0.1,1,10,100,1000,10000,1000000},
  yticklabels={$0.1$,$1$,$10$,$10^2$,$10^3$,$10^4$,$10^6$},
  xtick={0.1,0.2,0.3,0.4,0.5,0.6,0.7,0.8},
  tick label style={font=\scriptsize}, label style={font=\small},
  legend style={at={(0.5,-0.24)},anchor=north,font=\scriptsize,draw=none,fill=none,
     /tikz/every even column/.append style={column sep=0.30cm}},
  legend columns=4, legend cell align=left, clip=false]
\fill[gray!18] (axis cs:0.04,0.05) rectangle (axis cs:0.167,1e7);
\draw[gray!70,line width=0.6pt] (axis cs:0.167,0.05) -- (axis cs:0.167,1e7);
\draw[black!45,densely dotted,line width=0.5pt] (axis cs:0.13957,0.05) -- (axis cs:0.13957,1e7);
\draw[black!70,densely dashed,line width=0.6pt] (axis cs:0.354,0.05) -- (axis cs:0.354,1e7);
\draw[black!70,densely dashed,line width=0.6pt] (axis cs:0.542,0.05) -- (axis cs:0.542,1e7);
\input{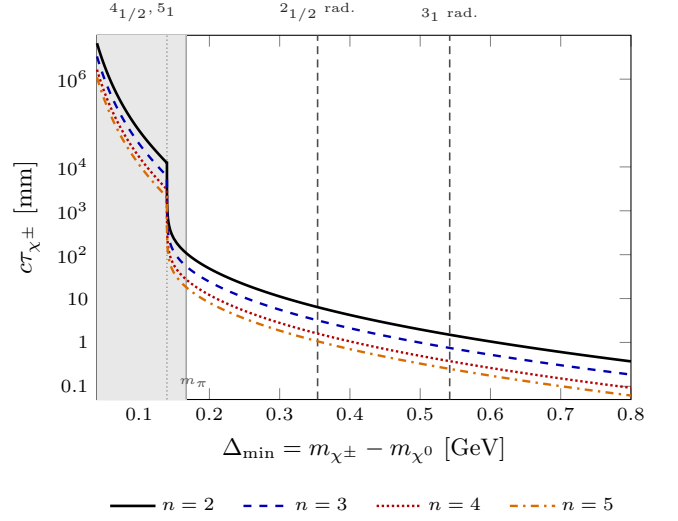}
\node[font=\tiny,anchor=west,black!70] at (axis cs:0.145,0.12) {$m_\pi$};
\node[font=\tiny,anchor=south,black!85] at (axis cs:0.105,1.35e7) {$4_{1/2},5_{1}$};
\node[font=\tiny,anchor=south,black!85] at (axis cs:0.354,1.35e7) {$2_{1/2}$ rad.};
\node[font=\tiny,anchor=south,black!85] at (axis cs:0.542,1.35e7) {$3_{1}$ rad.};
\end{axis}
\end{tikzpicture}
\caption{Proper decay length of the lightest charged partner as a function of its splitting from the DM. The curves use the one-pion width of Eq.~(\ref{eq:pionwidth}) and the heavy-to-heavy leptonic widths of Appendix~\ref{app:model}, with group-theory factor $j(j+1)-Y^2+Y$. They are continuous at the pion threshold (dotted vertical line). The dashed lines mark the radiative splittings of the doublet and $Y=1$ triplet. The grey region is the displayed part of the interval allowed by the central EFT bound for the quadruplet and quintuplet, Eq.~(\ref{eq:bound2}); it extends down to zero splitting. For these two multiplets the curves show the unmixed $Q=+1$ branch; the $Q=-1$ lifetimes at the same splitting are larger by factors $4/3$ and $3/2$. Charged mixing near the cap is not shown. A proper lifetime does not by itself identify a detector signature: the disappearing and surviving fractions depend on $p_T^\chi/m$ and the detector radii, as in Eq.~(\ref{eq:Ptrack}).}
\label{fig:ctau}
\end{figure}

The charged-partner lifetime depends strongly on its mass splitting from the DM (Fig.~\ref{fig:ctau}). For the larger multiplets, a useful upper bound follows from the leading splitting operators. We apply the charged-state sum rule and its lifetime implications discussed in Ref.~\cite{Bottaro2022} to the sub-MeV neutral splitting motivated by LZ.

Electroweak loops lift the state of charge $Q$ above the neutral one by~\cite{CFS,Bottaro2022}
\begin{equation}
 \Delta M_Q^{\rm EW}=\delta_g\left(Q^2+\frac{2YQ}{c_W}\right),
 \qquad \delta_g=167\pm4\,\mev ,
\label{eq:spectrum}
\end{equation}
which gives $0.35\,\gev$ for the doublet and $0.54\,\gev$ for the $Y=1$ triplet. For the two larger multiplets, however, Eq.~(\ref{eq:spectrum}) makes the state with $Q=-1$ lighter than the neutral one, so that the DM would be charged. These multiplets are viable only if an additional, ultraviolet contribution repairs the spectrum~\cite{Bottaro2022}, and for this reason their charged spectrum is often regarded as unknown.

We retain the leading dark-number-conserving splitting operator, together with the leading operator that generates $\dmz$, and assume that additional fields and higher-order operators give negligible corrections to this spectrum. For a fermionic multiplet the former operator can be written as
\begin{equation}
 \mathcal O_T=\frac{c_T}{\Lambda_T}\,(H^\dagger\sigma^aH)(\bar\chi\, T^a\chi),
\label{eq:OT}
\end{equation}
with a real dimensionless Wilson coefficient $c_T$. We use $H=(0,(v+h)/\sqrt2)^T$, $v=246\,\gev$, and include $+\mathcal O_T$ in the Lagrangian with mass term $-m\bar\chi\chi$. Thus $H^\dagger\sigma^3H=-v^2/2$ and the charged-neutral shift from this operator is $\Delta_Q^T=c_Tv^2Q/(2\Lambda_T)$. For a scalar multiplet we define $V_T=\lambda_T(H^\dagger\sigma^aH)(\chi^\dagger T^a\chi)$; its mass-squared shift gives $\Delta_Q^T=-\lambda_Tv^2Q/(4m)$ to first order in the splitting divided by the multiplet mass. Here $T^a$ are the isospin generators, $j=(n-1)/2$, and $T_3=Q-Y$. These operators change the term linear in $Q$ in Eq.~(\ref{eq:spectrum}). Let $M_0$ be the mean neutral mass, so that the DM mass is $M_0-\dmz/2$. The diagonal charged splittings are then $\delta_gQ^2+bQ+\dmz/2$, where $b$ includes the radiative and tree-level linear terms.

The neutral-splitting operator also mixes $\chi_{+1}$ with $\chi_{-1}^c$. Both fields in this basis have electric charge $+1$; their antiparticles form the charge-$-1$ sector. To leading order in the small splittings, the charged mass matrix relative to the DM mass is~\cite{Bottaro2022}
\begin{equation}
 \mathcal D_1=\left(\delta_g+\frac{\dmz}{2}\right)\mathbf 1
 +\begin{pmatrix}b&\eta\\\eta^*&-b\end{pmatrix},
 \qquad \eta=\mathcal O(\dmz).
\label{eq:chargedmatrix}
\end{equation}
Denote its eigenvalues by $\Delta_1\leq\Delta_2$. Their sum is its trace,
\begin{equation}
 \Delta_1+\Delta_2=2\delta_g+\dmz .
\label{eq:bound}
\end{equation}
Thus charged mixing preserves the sum rule within this operator truncation. Requiring a neutral ground state gives $\Delta_1>0$, and the lightest charged splitting obeys
\begin{equation}
 0<\dmin\ \le\ \delta_g+\tfrac12\dmz\simeq0.17\,\gev .
\label{eq:bound2}
\end{equation}
More precisely, $\dmin=\delta_g+\dmz/2-\sqrt{b^2+|\eta|^2}$, so mixing can only lower this upper bound. The small LZ-motivated $\dmz$ makes the bound close to the wino splitting. This is a result within the stated effective field theory (EFT), not a statement about arbitrary ultraviolet completions. For example, the dimension-seven fermionic operator
\begin{equation}
 \frac{c_7}{\Lambda^3}(H^\dagger\sigma^aH)(H^\dagger\sigma^bH)
 \bar\chi\{T^a,T^b\}\chi
\label{eq:highercharged}
\end{equation}
produces a term quadratic in $Q$ after symmetry breaking and can change the trace in Eq.~(\ref{eq:bound}). Its scalar analogue is dimension six, with $\bar\chi\{T^a,T^b\}\chi/\Lambda^3$ replaced by $\chi^\dagger\{T^a,T^b\}\chi/\Lambda^2$. Such contributions must be suppressed for the numerical cap to apply; a small neutral splitting alone does not guarantee their suppression.

Away from appreciable charged mixing, Eq.~(\ref{eq:pionwidth}) is multiplied by $C_Q=j(j+1)-Y^2+QY$. The $Q=+1,-1$ factors are $4,3$ for the quadruplet and $6,4$ for the quintuplet. Using the larger factor, the upper bound on the charged–neutral mass splitting evaluated at the central inputs and the leptonic widths of Appendix~\ref{app:model} gives the lifetime estimates
\begin{equation}
 c\tau_{\chi^\pm}\gtrsim27\,\mathrm{mm}\ \big[(4,\tfrac12)\big],
 \quad
 c\tau_{\chi^\pm}\gtrsim18\,\mathrm{mm}\ \big[(5,1)\big].
\label{eq:ctaubound}
\end{equation}
As the splitting decreases, the decay length grows rapidly but continuously through the pion threshold (Fig.~\ref{fig:ctau}). A proper lifetime of several meters can still yield a decay inside the tracker when the transverse boost is modest. The relative acceptance of disappearing-track and stable charged-particle searches is therefore set by the boosted decay length and detector geometry, rather than by a fixed splitting threshold (Sec.~\ref{sec:coverage}).

Within this EFT the bound applies whenever the multiplet contains both $Q=+1$ and $Q=-1$, that is, for $|Y|\le j-1$. The doublet and the $Y=1$ triplet have maximal hypercharge, $Y=j$, and no analogous two-state trace bound. For the larger multiplets the bound limits a splitting, rather than fixing a unique lifetime or track acceptance. Both singly charged mass eigenstates can be long-lived near the upper endpoint, where their splitting is of order $\dmz$ and mixing matters. Appendix~\ref{app:model} specifies how this limits the single-branch FCC-hh estimate.

\subsection{Beyond the minimal limit}
\label{sec:beyondminimal}

For fermionic doublets and triplets the radiative charged splitting is a prediction only when $|c_T|v^2/(2\Lambda_T)$ is small. In the normalization of Eq.~(\ref{eq:OT}), $|c_T|=1$ and $\Lambda_T=10^3\,\tev$ give a shift of $30\,\mev$ for $Q=1$. The neutral splitting does not determine this independent coefficient or scale.

The leading dark-number-violating operators have different dimensions for different hypercharges and spins~\cite{Bottaro2022}. They contain two dark fields and $4Y$ Higgs fields, and have dimension $4Y+3$ for fermions and $4Y+2$ for scalars. Their induced linear neutral splittings scale as
\begin{align}
 \dmz^{F}&\sim k_F\frac{v^{4Y}}{\Lambda_\delta^{4Y-1}},\nonumber\\
 \dmz^{S}&\sim k_S\frac{v^{4Y}}{m\Lambda_\delta^{4Y-2}},
\label{eq:neutralEFT}
\end{align}
where $k_F$ and $k_S$ include the Wilson coefficients and representation-dependent factors. For a fermion with $Y=1/2$ the operator is dimension five, so an effective coefficient of order unity and $\dmz\sim0.4\,\mev$ give a scale of order $10^8\,\gev$. For a fermion with $Y=1$ it is dimension seven and the scale instead varies as $(k_Fv^4/\dmz)^{1/3}$. For scalars, $Y=1/2$ permits a renormalizable quartic and fixes no ultraviolet scale; $Y=1$ requires a dimension-six operator with $\dmz\propto v^4/(m\Lambda_\delta^2)$. Thus $v^2/\dmz$ is not a common cutoff for the multiplets.

For a scalar doublet $\Phi=(H^+,(H^0+iA^0)/\sqrt2)^T$, define the relevant potential by~\cite{ArinaLingTytgat}
\begin{align}
 V\supset{}&m_2^2\Phi^\dagger\Phi
 +\lambda_3(H^\dagger H)(\Phi^\dagger\Phi)\nonumber\\
 &+\lambda_4|H^\dagger\Phi|^2
 +\left[\frac{\lambda_5}{2}(H^\dagger\Phi)^2+\mathrm{h.c.}\right].
\label{eq:IDMpotential}
\end{align}
For real $\lambda_5<0$, $H^0$ is the lighter neutral state. At tree level, $m_{H^+}^2-m_{H^0}^2=-(\lambda_4+\lambda_5)v^2/2$, giving
\begin{equation}
 \dmp^{\rm tree}\simeq-\frac{(\lambda_4+\lambda_5)v^2}{4m},
\label{eq:scalarsplit}
\end{equation}
while $\dmz\simeq-\lambda_5v^2/(2m)$. Equation~(\ref{eq:scalarsplit}) uses inert-doublet notation; it is not a generic formula for all scalar multiplets. For the $580\,\gev$ doublet, a charged shift below $10\,\mev$ requires $|\lambda_4+\lambda_5|\lesssim4\times10^{-4}$. For other representations we use the $\lambda_T$ convention defined above, together with their neutral-splitting operators.

Varying these independent splitting interactions gives the lifetime range in Fig.~\ref{fig:ctau}. Millimeter lifetimes motivate short-track reconstruction, while longer lifetimes can populate both disappearing-track and stable charged-particle searches. The transition depends on the transverse boost and detector geometry; closing the pion channel does not remove disappearing tracks. Searches using inclusive recoil or virtual effects therefore complement charged-track searches across this range.

Supersymmetry provides an instructive example. If the gauginos are very heavy, as the small neutral splitting suggests, they decouple from the charged splitting and the Higgsino is indistinguishable from the minimal doublet. If instead they are at the TeV scale and the neutral splitting is small because of a tuned cancellation, the charged splitting changes substantially, in a direction that depends on the sign of $M_2\mu$ and on $\tan\beta$: the chargino can become prompt, or long-lived enough to be visible at the LHC. The details are given in Appendix~\ref{app:susy}.

\subsection{What follows for the HL-LHC}

We can now assess the class as a whole. For the doublets the situation is that of Sec.~\ref{sec:lhc}; the scalar doublet is lighter than the Higgsino, but a pair of scalars is produced in a $p$ wave and its cross section is about ten times smaller than that of a fermion of the same mass, so it is equally out of reach. For all heavier candidates the obstacle is more basic. At $14\,\tev$, producing the heavier thermal multiplets requires increasingly large parton momentum fractions, strongly suppressing their rates; a quantitative reach depends on the representation and selection. The long-lived charged states motivated by the EFT bound in Eq.~(\ref{eq:bound2}) do not help if the particles are not produced in the first place.

The HL-LHC entries in Table~\ref{tab:multiplets} summarize the benchmark spectra considered here, including radiative doublet splittings. They do not establish exclusions or absence of sensitivity for every allowed charged spectrum. Longer-lived doublets, particularly the lighter scalar, require dedicated disappearing-track and stable charged-particle projections. The current fermionic limit of approximately $720\,\gev$ observed and $840\,\gev$ expected near $1\,\mathrm{ns}$~\cite{ATLAS2026} is a useful reference, but it is not a corresponding HL-LHC projection.

\subsection{Inelastic dark matter without electroweak charge}
\label{sec:nonEW}

Alternatives to dominant $Z$ exchange include dark photons~\cite{Yamashita2026,deLimaExo2026,Zhu2026}, additional $Z'$ bosons~\cite{HMLee2026,DuHuangXie2026,OkadaSeto2026,KumarPraj2026}, transition dipoles~\cite{ChangWeinerYavin,HeDipole2026,Asadi2026}, and axion portals~\cite{Unwin2026,YuanALP2026,AnAxion2026}. Their direct-detection normalization does not generally fix the collider production rate. For example, a heavy vector mediator with universal quark coupling $g_q$ and dark transition coupling $g_\chi$ gives a low-energy coefficient $C_q=g_qg_\chi/M_V^2$. At fixed dark mass, splitting and halo assumptions, the recoil rate constrains $|C_q|$, while collider production depends separately on the mediator mass, width, couplings and branching fractions~\cite{LHCDMWG}. Near an on-shell mediator, retaining only the contact coefficient is inadequate.

An illustrative translation discussed in Ref.~\cite{Jeon2026} uses a vector mediator, $m_\chi=1\,\tev$, $\delta=300\,\kev$, $g_q=0.25$ for each quark flavor, $g_\chi=1$, and no lepton coupling, following the simplified-model convention of Ref.~\cite{LHCDMWG}. The quoted $|C_q|\simeq(2$--$9)\times10^{-8}\,\gev^{-2}$ corresponds to $M_V\simeq1.67$--$3.54\,\tev$ for those couplings. This is a benchmark-dependent translation, not a mediator-mass prediction from the LZ event alone. The interval straddles the invisible pair threshold $M_V\simeq2m_\chi$: on-shell invisible decays are closed below it, whereas above it their branching fraction must be included in the recoil-search prediction. Dijet and, when allowed, dilepton searches probe visible decays; monojet and other missing-momentum searches probe dark-state production. A concrete leptophobic completion with a different charge assignment is studied in Ref.~\cite{DuHuangXie2026}.

Collider sensitivity in these models requires a scan over the mediator parameters and cannot be dismissed from the direct-detection normalization. The electroweak benchmarks considered here require no additional mediator: the relevant interactions are the Standard Model gauge interactions, with the independent portal couplings neglected by assumption.

\section{FCC-\lowercase{hh}}
\label{sec:fcc}

A $100\,\tev$ proton collider addresses both obstacles identified above. The higher energy raises the electroweak production rate of TeV-scale states by orders of magnitude, and a compact, fast inner tracker can reconstruct charged tracks of a few centimeters even under very high pileup. We first recall what has been established for the doublet and then extend the analysis to the other candidates, for which no dedicated studies exist.

\subsection{Higgsino doublet studies}
\label{sec:fccdoublet}

Three independent search strategies have been studied for the Higgsino at $100\,\tev$ with $30\,\abinv$.

\emph{Disappearing tracks.} A detailed detector-level study, including pileup, fake tracks, alternative tracker layouts and hit timing, finds that a Higgsino with $c\tau=7\,\mathrm{mm}$ can be discovered at $5\sigma$ up to $1.1$--$1.2\,\tev$ if the innermost tracker layers are moved close to the beam~\cite{Saito2019}. Extrapolations to the lower collision energies now under discussion give $5\sigma$ reaches of $1.0$--$1.1\,\tev$ at $80$--$84\,\tev$~\cite{FCCFSR2025,FCCPED2025}, so that the thermal mass sits at the edge of the discovery region. An independent study quotes a $95\%$~CL exclusion reach of $1.1$--$1.5\,\tev$, depending on the background assumptions~\cite{HanMukhoWang}. These numbers depend strongly on the tracker geometry~\cite{Saito2019,FCCPhysics2019}. This makes the LZ-motivated benchmark particularly informative: if the detector were to be able to validate the LZ excess, concrete design requirements to reconstruct centimeter-long tracks would have to be derived.

\emph{Monojet.} A search for missing momentum recoiling against a hard jet does not rely on the charged track at all. Its $95\%$~CL reach for the Higgsino is $1.4\,\tev$ if the background can be controlled at the level of $1\%$, and degrades quickly for larger systematic uncertainties~\cite{HanMukhoWang}.

\emph{Drell--Yan precision.} A heavy multiplet modifies the propagators of the electroweak gauge bosons and thereby distorts the invariant-mass and transverse-mass spectra of high-energy lepton pairs. A shape analysis of these spectra is sensitive to a Higgsino of up to $1.7\,\tev$ at the $2\sigma$ level, and would see the thermal Higgsino with a significance of $3.5\sigma$~\cite{ACEM}. This probe is also independent of the charged lifetime, and it measures the gauge quantum numbers directly.

The fermionic doublet is thus covered at $95\%$~CL by three complementary channels, two of which are insensitive to the charged spectrum. This matters because of the freedom discussed in Sec.~\ref{sec:beyondminimal}: even if the charged state is prompt and leaves no track, FCC-hh retains sensitivity to the thermal doublet.

\subsection{Extending the track search to the whole class}
\label{sec:coverage}

For the remaining candidates of Table~\ref{tab:multiplets} no dedicated FCC-hh disappearing-track study exists. A full simulation of each is beyond the scope of this paper, and it is also not necessary for our purpose. The candidates differ from the wino and the Higgsino, for which a complete analysis has been published~\cite{Saito2019}, in four respects relevant to our rescaling: their mass, their spin, their gauge quantum numbers, and the lifetime of their charged partner. The first three determine the production rate through the partonic cross sections and parton luminosities; the fourth determines the probability that the track is long enough to be reconstructed. All four enter the signal yield in a calculable way, so the published analysis can be transferred to the other multiplets by rescaling its signal, while its selections and background estimates are kept unchanged.

\emph{The model.} We describe the signal by the production of a pair of multiplet members together with a hard jet from initial-state radiation, which provides the missing transverse momentum required by the selection and boosts the pair. The yield is proportional to
\begin{equation}
 S\propto\sum_{\rm ch}g_{\rm ch}^2\!\int\! dM^2\,\hat\sigma(M)\!\int\!\frac{dp_\mathrm{T}}{p_\mathrm{T}}\,
 \mathcal L_{q\bar q}\big(\hat s\big)\,P_{\rm track},
\label{eq:yield}
\end{equation}
where $M$ is the invariant mass of the pair, $\hat\sigma(M)$ contains the velocity dependence of fermion or scalar pair production, $p_\mathrm{T}$ is the transverse momentum of the jet, integrated above the missing-momentum threshold of the analysis, and $\mathcal L_{q\bar q}(\hat s)$ is the quark--antiquark luminosity at the smallest partonic energy $\hat s(M,p_\mathrm{T})$ compatible with $M$ and $p_\mathrm{T}$. The couplings $g_{\rm ch}$ of all charged- and neutral-current channels follow from $n$ and $Y$. The last factor,
\begin{equation}
 P_{\rm track}=e^{-r_{\min}/\lambda_\mathrm{T}}-e^{-r_{\max}/\lambda_\mathrm{T}},
 \qquad \lambda_T=\frac{p_\mathrm{T}^{\chi}}{m}\,c\tau ,
\label{eq:Ptrack}
\end{equation}
is the probability that a charged state with transverse momentum $p_\mathrm{T}^\chi$ decays between the outermost tracker layer required by the analysis, $r_{\min}=150\,\mathrm{mm}$ for the five-hit tracks of the optimized layout of Ref.~\cite{Saito2019}, and the radius $r_{\max}\simeq1.5\,\mathrm{m}$ beyond which the track no longer disappears. We retain channels with one of the unmixed singly charged states, including prompt cascades from states with $|Q|>1$. For the larger multiplets we evaluate each possible assignment of the lightest branch with its own production weights and lifetime. The second singly charged state can also be long-lived. Appendix~\ref{app:model} explains when omitting its contribution gives a lower yield and where charged mixing invalidates this interpretation.

\emph{Validation.} The model has no free parameters apart from its overall normalization, which allows several nontrivial tests against published results. It reproduces the ratio of Higgsino to wino production cross sections and the mass dependence of the cross section between $1$ and $10\,\tev$ at next-to-leading order~\cite{AHHMW} to within $15\%$. It reproduces the dependence of the signal acceptance on the tracker layout reported in Ref.~\cite{Saito2019}, for both the wino and the Higgsino, to within $10\%$. These tests concern production rates and relative acceptances. We do not claim agreement with the published discovery significances: applying the same $30\%$ background uncertainty used for our projections changes them substantially (Appendix~\ref{app:model}). A further limitation is the relative normalization of the two signal regions of Ref.~\cite{Saito2019}, which use different missing-momentum thresholds: the Higgsino yield predicted from the wino one is too small by a factor of $3.5$. We therefore quote all results as a band, whose lower edge is normalized to the published wino yield and whose upper edge is normalized to the published Higgsino yield.

\emph{Criterion.} We ask whether FCC-hh can \emph{exclude} the benchmark at $95\%$~CL under the background-only hypothesis, rather than whether it can discover it at $5\sigma$, for two reasons: First, the HL-LHC limits with which we compare are $95\%$~CL exclusion limits. Second, the question raised by a direct-detection signal is whether its interpretation can be tested, and a hypothesis is tested once an experiment is sensitive to it. We therefore quote the expected upper limit $\mu_{95}$ on the signal strength $\mu$, defined so that the candidate itself corresponds to $\mu=1$; a candidate is covered when $\mu_{95}<1$. We also give the expected discovery significance $Z_{\rm disc}$, so that the reader can compare exclusion and discovery sensitivity. Expected limits are computed on a background-only Asimov dataset and the significance on a signal-plus-background one, in both cases with the asymptotic formulae for a counting experiment~\cite{Cowan2011,KumarMartin} and a $30\%$ uncertainty on the background. For each candidate we evaluate the two signal regions of Ref.~\cite{Saito2019}, which require missing transverse momentum above $2$ and $4\,\tev$ and expect $42.6$ and $1.6$ background events, and use the more sensitive one; exclusion needs $21.5$ and $3.1$ signal events, respectively, while $5\sigma$ discovery with the same background uncertainty needs $111.5$ and $11.9$ events. For all candidates the low-background region is the better choice, because the electroweak signals are small and a low background matters more than a high efficiency.

\emph{Continuous transition to long tracks.} The probability in Eq.~(\ref{eq:Ptrack}) applies on both sides of the pion threshold. The pion width vanishes continuously as $\dmin\to m_\pi^+$, while the leptonic widths remain finite, so neither the lifetime nor the track acceptance has a discontinuity. The probability to survive beyond a transverse radius $R$ is
\begin{equation}
 P_{\rm surv}(R)=e^{-R/\lambda_T} .
\label{eq:Psurvive}
\end{equation}
For example, a quintuplet with the $Q=+1$ branch lightest and $\dmin=139\,\mev$ has $c\tau\simeq2.10\,\mathrm m$. With $p_T^\chi/m=0.3$, $\lambda_T\simeq0.63\,\mathrm m$, Eq.~(\ref{eq:Ptrack}) gives $P_{\rm track}\simeq0.696$, while only $9\%$ survive beyond $1.5\,\mathrm m$. This is close to the maximum geometrical acceptance, attained at $\lambda_T=(r_{\max}-r_{\min})/\ln(r_{\max}/r_{\min})\simeq0.586\,\mathrm m$. We therefore scan the charged splitting continuously through the pion and lepton thresholds and retain the momentum-dependent acceptance inside the yield integral. At very long lifetimes $P_{\rm track}\simeq(r_{\max}-r_{\min})/\lambda_T$ decreases, producing a lower as well as an upper endpoint of the covered interval. Surviving tracks can instead enter searches for heavy stable charged particles, but Eq.~(\ref{eq:Psurvive}) alone is not their reconstruction efficiency or exclusion reach. Such searches require separate triggers, ionization and timing efficiencies, and background estimates; we do not assign them a sensitivity here.

\begin{table}[t]
\caption{Expected sensitivity of the FCC-hh disappearing-track search ($100\,\tev$, $30\,\abinv$) to the electroweak candidates. $N_{\rm sig}$ is the expected number of signal events in the more sensitive of the two signal regions of Ref.~\cite{Saito2019}; $\mu_{95}$ is the expected upper limit on the signal strength, so that the candidate is covered when it falls below unity, and $Z_{\rm disc}$ is the expected discovery significance. Ranges are ordered numerically: the conservative normalization gives the lower $N_{\rm sig}$ and $Z_{\rm disc}$ but the upper $\mu_{95}$. For the quadruplets and quintuplets, each row gives the smaller retained single-branch yield and its corresponding lifetime. The $167\,\mev$ rows extrapolate the gauge-basis calculation into the charged-mixing region (Appendix~\ref{app:model}). Radiative benchmarks are shown where defined, together with smaller splittings; the $139\,\mev$ rows explicitly lie below the pion threshold. Lifetimes and both normalization endpoints use the same continuous decay-width model. Entries marked $\gg1$ are far from any sensitivity.}
\label{tab:coverage}
\footnotesize
\begin{ruledtabular}
\setlength{\tabcolsep}{1.5pt}
\begin{tabular}{lccccc}
Candidate & $\dmin$ & $c\tau$ & $N_{\rm sig}$ & $\mu_{95}$ & $Z_{\rm disc}$ \\
 & [GeV] & [mm] & & & \\
\hline
$(2,\tfrac12)$ scalar & $0.354$ & $6.4$ & $9.4$--$33$ & $0.094$--$0.33$ & $4.2$--$9.8$ \\
$(2,\tfrac12)$ fermion & $0.354$ & $6.4$ & $2.5$--$8.7$ & $0.36$--$1.3$ & $1.5$--$4$ \\
$(3,1)$ scalar & $0.542$ & $0.75$ & ${\sim}0$ & ${\gg}1$ & ${\sim}0$ \\
 & $0.200$ & $24$ & $16$--$57$ & $0.055$--$0.2$ & $6.1$--$14$ \\
$(3,1)$ fermion & $0.542$ & $0.75$ & ${\sim}0$ & ${\gg}1$ & ${\sim}0$ \\
 & $0.200$ & $24$ & $25$--$87$ & $0.036$--$0.13$ & $8.2$--$17$ \\
$(4,\tfrac12)$ fermion & $0.167$ & $27$ & $3.7$--$13$ & $0.24$--$0.85$ & $2.1$--$5.3$ \\
 & $0.150$ & $75$ & $30$--$106$ & $0.03$--$0.1$ & $9.2$--$19$ \\
 & $0.139$ & $4.20\!\times\!10^3$ & $70$--$246$ & $0.013$--$0.045$ & $15$--$31$ \\
$(4,\tfrac12)$ scalar & $0.167$ & $27$ & $0.43$--$1.5$ & $2.1$--$7.3$ & $0.3$--$0.98$ \\
 & $0.145$ & $111$ & $4.4$--$16$ & $0.2$--$0.71$ & $2.4$--$6$ \\
 & $0.139$ & $4.20\!\times\!10^3$ & $4.2$--$15$ & $0.21$--$0.75$ & $2.3$--$5.8$ \\
$(5,1)$ fermion & $0.167$ & $18$ & ${\sim}0$ & ${\gg}1$ & ${\sim}0$ \\
 & $0.141$ & $168$ & $2.3$--$8.2$ & $0.38$--$1.3$ & $1.4$--$3.8$ \\
 & $0.139$ & $3.15\!\times\!10^3$ & $4.3$--$15$ & $0.21$--$0.73$ & $2.3$--$5.9$ \\
$(5,1)$ scalar & $0.167$ & $18$ & ${\sim}0$ & ${\gg}1$ & ${\sim}0$ \\
 & $0.141$ & $168$ & $0.058$--$0.2$ & $15$--$54$ & $0.042$--$0.15$ \\
 & $0.139$ & $3.15\!\times\!10^3$ & $0.08$--$0.28$ & $11$--$39$ & $0.058$--$0.2$ \\
\end{tabular}
\end{ruledtabular}
\end{table}

\emph{Results.} Table~\ref{tab:coverage} gives the expected sensitivity at representative splittings for each candidate, and Table~\ref{tab:windows} translates it into the range of charged splittings that the search covers. Several features deserve comment.

The scalar doublet is covered at its central radiative splitting in both normalizations. For the fermionic doublet our own estimate gives $\mu_{95}=0.36$--$1.27$: the optimistic normalization covers it, while the conservative one does not. The green entry in Fig.~\ref{fig:coverage} instead adopts the published $95\%$ CL reach of $1.1$--$1.5\,\tev$ for its $1.08\,\tev$ thermal mass. Our rescaling is therefore a marginal cross-check, not an independent exclusion throughout its normalization band. The scalar-doublet result comes from our rescaling. Its production is suppressed near threshold because a pair of scalars is produced in a $p$ wave, so that the cross section rises only as $\beta^3$ with the velocity $\beta$ of the pair rather than as $\beta$, but this is more than compensated by its lower mass, which raises both the parton luminosity and the boost that carries the charged state into the tracker.

The triplets are covered over finite intervals of charged splitting: approximately $80$--$228\,\mev$ for the scalar and $56$--$229\,\mev$ for the fermion with the conservative normalization. At their radiative splitting of $0.54\,\gev$ the lifetime is below a millimeter and the model yields negligible track acceptance. Reducing the splitting to the upper end of the covered interval requires a negative custodial-breaking contribution. For the fermion, the corresponding scale is of order $\Lambda_T/|c_T|\simeq100\,\tev$ in the convention of Eq.~(\ref{eq:OT}). At the lower end, the leptonic decay becomes so slow that the probability to disappear inside the tracker is too small. Since the triplets have maximal hypercharge, nothing forces their splitting into these intervals, so disappearing-track searches alone cannot test these candidates over the full range of allowed splittings; we return to lifetime-independent searches in Sec.~\ref{sec:indirect}.

The single-branch estimate for the fermionic quadruplet gives a window of approximately $70$--$167\,\mev$ with the conservative normalization, or $55$--$167\,\mev$ with the optimistic one. Both intervals extend substantially below the pion threshold. They occupy $58\%$ and $67\%$, respectively, of the full positivity-allowed interval $0<\dmin\leq167\,\mev$, using the central spectral bound of Sec.~\ref{sec:spectrum}. These percentages are lengths in $\dmin$, not probabilities over ultraviolet parameters. At the formal upper endpoint the unmixed $Q=+1$ branch gives the smaller yield, $3.7$--$13$ events and $\mu_{95}=0.24$--$0.85$; at $139\,\mev$ the $Q=-1$ branch instead gives the smaller yield, about $70$--$246$ events. The large production index and the spectral bound make a broad interval accessible, but do not establish coverage of arbitrarily small splittings. The $167\,\mev$ endpoint lies in the region where the gauge-branch approximation can fail; its coverage requires the mixed-state treatment described in Appendix~\ref{app:model}. The spectral-bound uncertainty also affects this endpoint.

The remaining three candidates are more difficult. The scalar quadruplet is covered over approximately $130$--$148\,\mev$ in the conservative normalization and $98$--$161\,\mev$ in the optimistic one. The fermionic quintuplet is covered over $125$--$140\,\mev$ or $91$--$144\,\mev$, respectively; most of the newly included interval lies below the pion threshold. At $139\,\mev$ its conservative branch has $c\tau\simeq3.15\,\mathrm m$ and gives $4.3$--$15$ events, or $\mu_{95}=0.21$--$0.73$, so even the lower normalization covers this point. This branch is $Q=-1$, whereas the geometrical example above used $Q=+1$. The scalar quintuplet at $11.5\,\tev$ remains outside the disappearing-track reach throughout the scan. At still smaller splittings the charged states increasingly survive the tracker; their possible coverage by stable charged-particle searches is unquantified here.

\begin{table}[t]
\caption{Charged-splitting intervals of the FCC-hh single-branch disappearing-track estimate at $95\%$~CL, including the continuous transition through $m_\pi$. Central-input endpoints are rounded to the nearest MeV; this rounding is not an uncertainty estimate. Table~\ref{tab:inputvariation} gives their variation with the quoted thermal masses and radiative splitting. The two columns use the conservative and optimistic normalizations. For the quadruplets and quintuplets, the last column gives the corresponding interval lengths divided by the full central EFT range, $0<\dmin\leq167\,\mev$; these fractions do not represent probabilities for model parameters. The $167\,\mev$ endpoint is a gauge-basis extrapolation requiring a mixed-state calculation. For doublets and triplets the status refers to the radiative splitting. Stable charged-particle searches are not included.}
\label{tab:windows}
\footnotesize
\begin{ruledtabular}
\begin{tabular}{@{}l@{\quad}c@{\quad}c@{\quad}l@{}}
Candidate & \multicolumn{2}{c}{$\dmin$ [MeV]} & fraction/status \\
& conservative & optimistic & \\
\hline
$(2,\tfrac12)$ scalar & $88$--$385$ & $68$--$419$ & covered \\
$(2,\tfrac12)$ fermion & $64$--$349$ & $50$--$376$ & at the edge \\
$(3,1)$ scalar & $80$--$228$ & $62$--$248$ & not radiative \\
$(3,1)$ fermion & $56$--$229$ & $43$--$246$ & not radiative \\
$(4,\tfrac12)$ fermion & $70$--$167$ & $55$--$167$ & $58\%$ ($67\%$) \\
$(4,\tfrac12)$ scalar & $130$--$148$ & $98$--$161$ & $11\%$ ($38\%$) \\
$(5,1)$ fermion & $125$--$140$ & $91$--$144$ & $9\%$ ($32\%$) \\
$(5,1)$ scalar & none & none & not covered \\
\end{tabular}
\end{ruledtabular}
\end{table}

\emph{Input dependence.} The quoted intervals use central thermal masses and $\delta_g=167\,\mev$. Independently varying each mass over the quoted lower, central and upper values and taking $\delta_g=163,167,171\,\mev$ gives the endpoint spreads in Table~\ref{tab:inputvariation}; Appendix~\ref{app:model} specifies the prescription. These variations are separate from the yield-normalization band. The scalar-doublet radiative benchmark remains covered, whereas the fermionic-doublet conservative limit ranges over $\mu_{95}=0.78$--$2.08$. For the fermionic quadruplet, even the gauge-basis estimate at the varied spectral cap gives $\mu_{95}=0.35$--$2.03$ with the conservative normalization, so coverage up to the cap is not uniform. The conservative interval fractions vary from $54\%$ to $60\%$ for the fermionic quadruplet, $6\%$ to $17\%$ for the scalar quadruplet, and approximately $0.09\%$ to $19\%$ for the fermionic quintuplet. At the upper quintuplet mass, $10.6\,\tev$, only a roughly $0.16\,\mev$ window just above $m_\pi$ remains in this approximation. Its width is smaller than the neglected neutral splitting, so it demonstrates the fragility of the central estimate rather than a controlled sub-MeV exclusion prediction. The scalar quintuplet has no track window for any of the sampled inputs.

We emphasize the status of these statements. Equation~(\ref{eq:yield}) is a rescaling model constrained by published production and acceptance checks. Its factor-$3.5$ band describes the two yield normalizations; it does not quantify all detector uncertainties. In extending the scan to long lifetimes, we retain the published efficiencies and backgrounds apart from the explicit geometrical decay probability. Whether longer tracks, surviving charged partners, and changes in reconstructed missing momentum preserve these selections requires a detector study. Tables~\ref{tab:coverage} and~\ref{tab:windows} should therefore be read as conditional estimates, especially near either endpoint of a covered interval. We neither impose a discontinuity at $m_\pi$ nor assume that stable charged-particle searches fill the remaining gaps.

\subsection{Lifetime-independent channels}
\label{sec:indirect}

Track searches fail when the charged state decays promptly, as it does for the triplets at their radiative splitting. The monojet and Drell--Yan channels of Sec.~\ref{sec:fccdoublet} do not have this limitation, so it is natural to ask what they can contribute for the other candidates.

For the monojet search we apply the same yield model without the track requirement and compare each candidate with the two published reach points of Ref.~\cite{HanMukhoWang}, a wino of $2.0\,\tev$ and a Higgsino of $1.4\,\tev$. The model predicts equal accepted rates for these two points to within $15\%$, as it should if it describes the mass and representation dependence correctly. The fermionic triplet at its thermal mass gives a rate equal to or slightly above these reference points and is therefore marginally covered, under the same optimistic assumption of $1\%$ background systematics. All other candidates fall short: the fermionic quadruplet by a factor of $1.4$--$2$, the scalar doublet and triplet by factors of $2.5$--$4$, and the scalar quadruplet and both quintuplets by more than an order of magnitude.

For the Drell--Yan channel the effect of a multiplet is governed by two coefficients, $C_1\propto nY^2$ and $C_2\propto n(n^2-1)$, which multiply its contributions to the hypercharge and $SU(2)_L$ gauge-boson self-energies~\cite{ACEM}. Well below the pair threshold the distortion scales as $(C_1+C_2)/m^2$. Normalizing to the fermionic doublet, this quantity is $1.2\,\tev^{-2}$ for the fermionic triplet at its thermal mass, well above the value of $0.7\,\tev^{-2}$ at which the published analysis reaches $2\sigma$ sensitivity for the doublet. We therefore expect the fermionic triplet to be covered by the Drell--Yan measurement irrespective of its charged lifetime. The scalar doublet is marginal by the same measure ($0.74\,\tev^{-2}$). The fermionic quadruplet ($0.52\,\tev^{-2}$) lies between the thresholds inferred from the published doublet and wino reaches, so the scaling argument does not settle it; since its pair threshold lies above the fitted mass range of the analysis, where only the smooth tail below threshold contributes, we do not consider it here. The remaining candidates fall short at $0.3\,\tev^{-2}$ or below. 

Combining published doublet studies with our estimates at central theory inputs, FCC-hh probes both radiative doublets and excludes the fermionic quadruplet at $95\%$~CL over approximately $70$--$167\,\mev$ in the single-branch estimate with the lower normalization and the endpoint qualification of Appendix~\ref{app:model}; it covers the fermionic triplet through the Drell--Yan measurement, and marginally through the monojet search, irrespective of its charged lifetime; and it covers the scalar triplet, the scalar quadruplet and the fermionic quintuplet over more limited intervals of charged splitting, including sub-pion values. Stable charged-particle searches could complement these finite windows but have not been evaluated here. Only the scalar quintuplet is beyond reach in every channel we have examined. This substantially extends the reach beyond the HL-LHC benchmark projections considered here.

\section{Muon collider}
\label{sec:muc}

A high-energy muon collider approaches the same targets from a different direction, and in several respects it is better matched to them than a hadron collider. We summarize why, and what the existing studies imply for the candidates of Table~\ref{tab:multiplets}.

\subsection{Why a muon collider is well suited}

Muons are elementary, so electroweak pair production can happen up to masses close to the kinematic limit $\sqrt s/2$.\footnote{At a hadron collider the quintuplets are limited by the collapse of the parton luminosity near $20\,\tev$.}  How close to the limit a search reaches depends on the lifetime: for a long-lived state such as the wino the disappearing-track reach is close to $\sqrt s/2$, whereas for the short-lived doublet it is about $0.3\sqrt s$~\cite{HanMuC,Capdevilla2021}.

The relatively clean environment allows searches for a single photon or a single $W$ boson recoiling against invisible particles. Their reach depends strongly on the systematic uncertainty achieved, and the projections below are quoted for uncertainties between zero and one per mille, with the value for one per cent given where it changes the conclusion~\cite{HanMuC,Bottaro2022}. These signatures count the production of \emph{any} member of the multiplet and do not depend on the charged lifetime. This removes the main source of uncertainty in the hadron-collider results: for the doublet and the triplet, whose charged splitting is not predicted (Sec.~\ref{sec:beyondminimal}), a muon collider can decide the question without knowing the spectrum.

For the larger multiplets the EFT bound in Eq.~(\ref{eq:bound2}) motivates charged-track searches, with acceptance determined by the lifetime and boost. Searches for disappearing and for long charged tracks at a muon collider have been studied including the beam-induced background, and they cover the thermal Higgsino and wino at $\sqrt s=10\,\tev$~\cite{Capdevilla2021}; with soft tracks the thermal Higgsino is within reach already at $3\,\tev$~\cite{CapdevillaSoft}.

At a hadron collider scalar pairs are produced near threshold, where their $p$-wave cross section is an order of magnitude below that of fermions. At a lepton collider the pair is produced at the full collision energy, and for $\sqrt s\gg2m$ the suppression is only the asymptotic factor of four.

\subsection{Coverage of the class}

The \emph{fermionic} multiplets of Table~\ref{tab:multiplets} have been studied at muon-collider energies between $3$ and $30\,\tev$, with an integrated luminosity of $10\,\abinv\times(\sqrt s/10\,\tev)^2$~\cite{Bottaro2022}. The doublet is reached by the soft-track search already at $3\,\tev$~\cite{CapdevillaSoft}; the mono-$W$ reach at $6\,\tev$, $1.0\,\tev$, falls just short of its thermal mass with that luminosity, and at $10\,\tev$ it is $1.5\,\tev$ ($1.3\,\tev$ for one per cent systematics), so the candidate is comfortably covered. The triplet is covered at $10\,\tev$, the baseline energy of the current design effort~\cite{IMCC2025,Accettura2023}, with a mono-$\gamma$ reach of $3.5$--$3.6\,\tev$ against a thermal mass of $2.85\,\tev$; this margin does require per-mille systematics, since for one per cent the reach falls to $2.7\,\tev$. At the radiative charged splitting, this is a candidate that FCC-hh addresses principally through indirect measurements. The quadruplet needs more than $10\,\tev$, its pair-production threshold being $9.6\,\tev$: the mono-$\gamma$ and mono-$W$ reaches grow from $3.8$--$4.1\,\tev$ at $10\,\tev$ to $5.2$--$5.5\,\tev$ at $14\,\tev$, above its $4.8\,\tev$ thermal mass, although for one per cent systematics the $14\,\tev$ reach drops to $4.3$--$4.4\,\tev$ and coverage then relies on the track searches. The quintuplet, with a threshold of $19.8\,\tev$, is covered by charged-track searches for $\sqrt s>20\,\tev$ and by all channels at $30\,\tev$~\cite{Bottaro2022}. Precision measurements of high-energy fermion pair production can extend the sensitivity beyond the pair-production threshold~\cite{FranceschiniZhao}, in analogy to the Drell--Yan analysis at FCC-hh. Scalar projections are given in Fig.~10 and Appendix~D of Ref.~\cite{Bottaro2022}. The one-track search reaches the thermal doublet at $6\,\tev$ with $4\,\abinv$ and the thermal quadruplet at $14\,\tev$ with $20\,\abinv$. The combined missing-invariant-mass search reaches the thermal triplet at $14\,\tev$ for sub-percent systematics. At $30\,\tev$ and $90\,\abinv$, the strongest scalar-quintuplet reach is about $11\,\tev$, short of its central thermal mass $11.5\,\tev$ but within the quoted $\pm0.8\,\tev$ mass variation. Its classification is therefore sensitive to the thermal-mass input. These track projections use the neutral splittings and charged spectra specified in that reference, including neutral splittings of several MeV for scalars. Their transfer to the smaller LZ-motivated neutral splittings requires a lifetime and spectrum recast; the recoil searches are insensitive to such small neutral splittings. We therefore quote the published scalar benchmarks explicitly and do not identify pair-production thresholds with exclusion reaches.

\begin{figure*}[t]
\centering
\begin{tikzpicture}[x=1.72cm,y=0.66cm,font=\small]
\definecolor{cyes}{RGB}{56,132,72}
\definecolor{cpart}{RGB}{226,163,42}
\definecolor{cno}{RGB}{192,72,62}
\definecolor{cna}{RGB}{165,165,165}
\newcommand{\cell}[4]{\fill[#3,rounded corners=1.4pt] (#1-0.43,#2-0.31) rectangle (#1+0.43,#2+0.31);
  \node[white,font=\scriptsize\bfseries] at (#1,#2) {#4};}
\newcommand{\Y}[2]{\cell{#1}{#2}{cyes}{$\checkmark$}}
\newcommand{\Pp}[2]{\cell{#1}{#2}{cpart}{$\sim$}}
\newcommand{\N}[2]{\cell{#1}{#2}{cno}{$\times$}}
\newcommand{\Q}[2]{\cell{#1}{#2}{cna}{?}}
\draw[black!55,line width=0.5pt] (0.57,2.05) -- (3.43,2.05);
\node[font=\scriptsize\bfseries] at (2,2.45) {FCC-hh\quad ($100\,\tev$, $30\,\abinv$)};
\draw[black!55,line width=0.5pt] (3.57,2.05) -- (6.43,2.05);
\node[font=\scriptsize\bfseries] at (5,2.45) {muon collider};
\foreach \c/\lab in {0/{HL-LHC$^a$}, 1/{disappearing\\tracks}, 2/{mono-$j$}, 3/{Drell--Yan}, 4/{$3\,\tev$}, 5/{$10\,\tev$}, 6/{$30\,\tev$}}
  \node[align=center,font=\scriptsize\bfseries] at (\c,1.3) {\lab};
\node[anchor=east,font=\scriptsize] at (-0.62,0)  {$(2,\tfrac12)$ scalar,\ \ $0.58\,\tev$};
\node[anchor=east,font=\scriptsize] at (-0.62,-1) {$(2,\tfrac12)$ fermion,\ \ $1.08\,\tev$};
\node[anchor=east,font=\scriptsize] at (-0.62,-2) {$(3,1)$ scalar,\ \ $2.1\,\tev$};
\node[anchor=east,font=\scriptsize] at (-0.62,-3) {$(3,1)$ fermion,\ \ $2.85\,\tev$};
\node[anchor=east,font=\scriptsize] at (-0.62,-4) {$(4,\tfrac12)$ fermion,\ \ $4.8\,\tev$};
\node[anchor=east,font=\scriptsize] at (-0.62,-5) {$(4,\tfrac12)$ scalar,\ \ $4.98\,\tev$};
\node[anchor=east,font=\scriptsize] at (-0.62,-6) {$(5,1)$ fermion,\ \ $9.9\,\tev$};
\node[anchor=east,font=\scriptsize] at (-0.62,-7) {$(5,1)$ scalar,\ \ $11.5\,\tev$};
\N{0}{0}\N{0}{-1}\N{0}{-2}\N{0}{-3}\N{0}{-4}\N{0}{-5}\N{0}{-6}\N{0}{-7}
\Y{1}{0}\cell{1}{-1}{cyes}{$\checkmark^{*}$}\Pp{1}{-2}\Pp{1}{-3}\cell{1}{-4}{cpart}{$58$--$67\%$}\cell{1}{-5}{cpart}{$11$--$38\%$}\cell{1}{-6}{cpart}{$9$--$32\%$}\N{1}{-7}
\N{2}{0}\Pp{2}{-1}\N{2}{-2}\Pp{2}{-3}\Pp{2}{-4}\N{2}{-5}\N{2}{-6}\N{2}{-7}
\Pp{3}{0}\Y{3}{-1}\N{3}{-2}\Y{3}{-3}\Pp{3}{-4}\N{3}{-5}\N{3}{-6}\N{3}{-7}
\Q{4}{0}\Y{4}{-1}\N{4}{-2}\N{4}{-3}\N{4}{-4}\N{4}{-5}\N{4}{-6}\N{4}{-7}
\Q{5}{0}\Y{5}{-1}\Q{5}{-2}\Y{5}{-3}\N{5}{-4}\Q{5}{-5}\N{5}{-6}\N{5}{-7}
\Q{6}{0}\Y{6}{-1}\Q{6}{-2}\Y{6}{-3}\Y{6}{-4}\Q{6}{-5}\Y{6}{-6}\N{6}{-7}
\node[anchor=west,font=\scriptsize] at (-2.9,-8.25)
 {\tikz\fill[cyes,rounded corners=1pt](0,0)rectangle(0.28,0.22); covered at $95\%$ CL \quad
  \tikz\fill[cpart,rounded corners=1pt](0,0)rectangle(0.28,0.22); conditional or marginal \quad
  \tikz\fill[cna,rounded corners=1pt](0,0)rectangle(0.28,0.22); no reach quoted at this energy \quad
  \tikz\fill[cno,rounded corners=1pt](0,0)rectangle(0.28,0.22); not covered};
\end{tikzpicture}
\caption{Collider coverage at central thermal masses, using $95\%$ CL exclusion. $^a$HL-LHC entries apply to the benchmark spectra, including radiative doublet splittings, not arbitrary charged lifetimes. FCC-hh combines published fermionic-doublet studies with our track and monojet rescaling and Drell--Yan estimate. $^*$The green fermionic-doublet track cell uses the published reach; our normalization band straddles $\mu_{95}=1$ (Table~\ref{tab:coverage}). Doublet tracks assume radiative splittings. Conditional cells depend on the charged spectrum, systematics, or marginal scaling estimates. Track percentages give interval lengths for the two yield normalizations relative to $0<\dmin\leq167\,\mev$, not parameter probabilities. They include sub-pion splittings but no stable charged-particle coverage. Input variations can change marginal windows (Table~\ref{tab:inputvariation}); the quadruplet upper endpoint requires charged mixing (Appendix~\ref{app:model}). Fermionic muon-collider entries follow Refs.~\cite{Bottaro2022,Capdevilla2021,CapdevillaSoft,HanMuC}, with $1$, $10$ and $90\,\abinv$ at $3$, $10$ and $30\,\tev$. Published scalar benchmarks are DT at $6\,\tev$ for the doublet and $14\,\tev$ for the quadruplet, and MIM at $14\,\tev$ for the triplet~\cite{Bottaro2022}; these energies are not columns here. Gray means no quoted reach at that energy, not kinematic exclusion: the scalar-quadruplet pair threshold is $9.96\pm0.50\,\tev$. The scalar-quintuplet $30\,\tev$ reach misses the central mass but overlaps its uncertainty range. Scalar DT results require a spectrum recast for LZ.}
\label{fig:coverage}
\end{figure*}

\subsection{Complementarity with FCC-hh and other machines}

Figure~\ref{fig:coverage} compares the energy-frontier machines. The two machines probe the class in different ways, with unrelated systematic limitations: tracks and loop corrections to Drell--Yan production in a hadronic environment on one side, recoil spectra and tracks in the presence of beam-induced backgrounds on the other.  A $10\,\tev$ muon collider settles the fermionic doublet and triplet regardless of their charged spectrum, a 30\,TeV muon collider has projected sensitivity to all fermionic benchmarks considered here. This includes the quintuplet, which our FCC-hh estimate reaches over approximately $125$--$140\,\mev$ with the conservative normalization, extending to $91$--$144\,\mev$ with the optimistic one, at central theory inputs. The mass variation can narrow this FCC-hh window substantially (Table~\ref{tab:inputvariation}). At the same time, FCC-hh covers the fermionic quadruplet over approximately $70$--$167\,\mev$ in the single-branch estimate with the lower normalization and the endpoint qualification of Appendix~\ref{app:model}, for which a $10\,\tev$ muon collider has too little energy. We point out that the FCC-hh-based track and Drell--Yan measurements determine the lifetime and the gauge quantum numbers, while muon collider-based recoil searches provide a handle on the mass and the production rate. In this sense a muon collider of sufficient energy provides a better coverage of the model class and depends less on the unknown charged spectrum.


Low-energy electron--positron colliders, defined here by $\sqrt s\leq500\,\gev$, cannot pair-produce any of these thermal benchmarks: even the lightest scalar doublet has a central threshold $2m_{\rm th}=1.16\,\tev$. This includes the sub-$0.4\,\tev$ programs of FCC-ee and CEPC. This kinematic statement concerns direct production; precision effects below threshold are a separate probe~\cite{Harigaya2015}, for which we do not assign a low-energy exclusion reach here.

Multi-TeV electron--positron machines are different. The current CLIC proposal includes a $1.5\,\tev$ stage, with extension towards $2\,\tev$, and an earlier design documents $3\,\tev$~\cite{CLIC2025}. The doublet pair thresholds are $1.16\,\tev$ for the scalar and $2.16\,\tev$ for the fermion, so a $1.5\,\tev$ stage opens the scalar channel and a $3\,\tev$ stage opens both. Published $3\,\tev$ CLIC projections include sensitivity to a thermal Higgsino through short charged tracks, conditional on the assumed background and tracking performance~\cite{CLIC2018}. The $3\,\tev$ option remains below the pair thresholds of the triplets and larger thermal multiplets, which are the principal reason to consider FCC-hh and higher-energy muon colliders.

\section{Caveats and outlook}
\label{sec:caveats}

Our collider conclusions depend on the interpretation of the keV recoil, the halo velocity distribution and the assumptions used to select the multiplets. Below we discuss these assumptions and their uncertainties, and the measurements needed to resolve them.

\emph{The LZ event rate.} The LZ result with one event has a global significance of $2.6\sigma$~\cite{LZ2026}. A background fluctuation, an unmodeled detector effect, or an origin that does not involve dark-matter scattering at all, such as the disappearance of neutrons bound in the xenon nucleus~\cite{AghaieStrumia2026,LeeTakahashi2026} or ordinary neutrino scattering~\cite{Chattaraj2026}, remain plausible explanations. Additional events would test the recoil shape and constrain the splitting and halo assumptions. They would not automatically determine a very heavy DM mass: for $m_\chi\gg m_N$, the reduced mass approaches $m_N$ and the kinematic dependence on $m_\chi$ remains weak. An independent mass measurement or stronger assumptions about the rate normalization would still be needed. Other target nuclei provide complementary tests of the kinematic edge~\cite{SuYangYang2026,BaerBargerAr2026,AhmedEdge2026,GeTitov2026,Palmisano2026,GuLi2026}.

\emph{Astrophysical and solar constraints.} Because the event must come from the fastest halo particles, the inferred splitting is unusually sensitive to the velocity distribution, and it can shift by $\mathcal O(100\,\kev)$~\cite{FanReece2026,GhoshChavezKelso2026}. Such shifts have negligible effects on inclusive production and on charged decays away from degeneracy; the charged mixing region near the spectral cap still depends on the neutral splitting. On the other hand, solar-capture limits and the empty LZ sideband discussed in Sec.~\ref{sec:noncollider} have more subtle implications. They apply in a qualitatively similar way to every candidate in Table~\ref{tab:multiplets}, since the capture cross section is fixed by the same hypercharge that fixes the LZ rate, but not with the same strength: the capture and annihilation rates, the approach to equilibrium and the neutrino spectra all depend on the mass and the representation, so that each candidate requires its own analysis~\cite{DiMauroMDM2026}. A joint treatment of the event, the sideband and the solar limits, with a common halo model and a fitted signal normalization, is the most important missing piece on the noncollider side.

\emph{Model dependence.} Our target class is defined by the assumptions in the Introduction: one approximately pure multiplet with an exact stabilizing $\mathbb Z_2$, dominant $Z$ exchange, a relic abundance accounting for all dark matter and set by gauge-dominated thermal freeze-out, and negligible independent portal or higher-order interactions beyond the retained splitting operators. Relaxing the scattering assumption leads to the mediator models of Sec.~\ref{sec:nonEW}, to boosted or exothermic dark matter~\cite{Alhazmi2026,Kannike2026,MahapatraPaul2026,deLimaExo2026,BaerBarger2026,Xing2026}, to dark-matter absorption~\cite{LouLu2026}, or to elastic interpretations, which do not in general require electroweak charged partners. Relaxing the relic-density assumption removes the stated mass prediction; lighter, subdominant multiplets would be easier to find. The restriction on additional interactions controls the charged spectrum of the doublet and triplet, which is why lifetime-independent searches are essential for them. Higher-order operators with additional even powers of charge can also change the larger-multiplet bound, as illustrated in Sec.~\ref{sec:spectrum}.

\emph{Next steps.} On the collider side, our FCC-hh estimates rescale a published analysis and should be replaced by dedicated studies. The most informative ones would be a calculation retaining both charged mass eigenstates and their cascades near degeneracy, a combined disappearing-track and stable charged-particle analysis across the full lifetime range, including the sub-pion region and the smallest splittings where our disappearing-track acceptance falls below sensitivity, a Drell--Yan analysis for the two triplets, and a recast of the existing scalar muon-collider projections~\cite{Bottaro2022} for the LZ-motivated neutral and charged spectra. In terms of detector design, the conclusion drawn from the doublet scenario is that the radius of the innermost tracking layers and the control of systematic uncertainties in the monojet and Drell--Yan measurements directly decide whether a thermal target is covered~\cite{Saito2019,FCCFSR2025}. At the HL-LHC, dedicated track triggers~\cite{CMSL1TDR} motivate further studies of long-lived doublets and lighter, non-thermal multiplets.

\section{Conclusions}
\label{sec:conc}

The high-energy nuclear recoil event reported by LZ does not constitute a discovery of dark matter, but it raises a well-defined question: if the event is due to inelastic dark matter carrying electroweak charge, how could this be established? We have addressed the collider side of this question.

The thermal Higgsino, the most discussed interpretation, cannot be tested at the HL-LHC. Its neutral states are invisible because their splitting is below the electron-pair threshold, and its charged partner is produced too rarely and decays too early for disappearing-track searches to reach its mass. These benchmark limitations follow from the electroweak spectrum rather than from supersymmetry itself. In the approximately pure, stable multiplets studied here, the halo-dependent recoil interpretation motivates a small neutral splitting, gauge quantum numbers determine the production interactions, and gauge-dominated thermal freeze-out gives masses between $0.58$ and $11.5\,\tev$. Charged lifetimes vary across the class and are treated separately. Applying the existing charged-state sum rule within the specified EFT, we find that a neutral ground state bounds the lightest charged splitting in the larger multiplets. This motivates long-lived-particle searches but does not establish a universal lifetime or acceptance for arbitrary ultraviolet completions.

These properties make the class a well-defined target for the energy frontier. Rescaling a published detector study with a validated model of the signal yield, and using an expected $95\%$~CL exclusion as the criterion, we find that FCC-hh probes the radiative doublets, adopting the published reach for the fermionic case, and at the central theory inputs covers the fermionic quadruplet over approximately $70$--$167\,\mev$ in the single-branch estimate with the lower normalization and the endpoint qualification of Appendix~\ref{app:model}, covers the fermionic triplet through a Drell--Yan measurement independent of the lifetime, and covers the scalar triplet, the scalar quadruplet and the fermionic quintuplet over more limited charged-splitting intervals that extend below the pion threshold; only the scalar quintuplet is beyond reach in every FCC-hh channel we examined. The thermal-mass and radiative-splitting variations in Table~\ref{tab:inputvariation} weaken marginal track claims, particularly the fermionic-quintuplet window. Muon-collider recoil searches provide sensitivity independent of the charged lifetime. Published projections at \(10\,\mathrm{TeV}\) cover the thermal fermionic doublet and triplet, while about $14$ and $30\,\tev$ are required for the fermionic quadruplet and quintuplet; published scalar projections reach the doublet and quadruplet through tracks at $6$ and $14\,\tev$, and the triplet through recoil searches at $14\,\tev$, subject to the spectrum and systematic assumptions discussed above, while the central thermal scalar quintuplet remains beyond the published $30\,\tev$ reach. FCC-hh and high-energy muon colliders offer complementary probes of the heavier thermal multiplets. Low-energy electron--positron machines with $\sqrt s\leq500\,\gev$ cannot directly produce these benchmarks, whereas $3\,\tev$ CLIC opens doublet production and has projected sensitivity to the thermal Higgsino.

If the LZ high-recoil signal grows with exposure and survives the solar and sideband constraints, testing the specified thermal electroweak interpretations will motivate a collider at the energy frontier beyond the LHC. The classification presented here specifies what such a machine has to deliver: sensitivity to electroweak states of several TeV, tracking of charged particles with decay lengths from centimeters to meters, and searches that remain sensitive when no track is left at all.

\begin{acknowledgments}
We acknowledge the use of LLMs (Claude Fable and GPT-6 Astra) for our studies. Fable was used to compile existing results from the literature, for extrapolations, and for creating a base version of the manuscript text. GPT-6 Astra was used to cross-check and refine Fable's initial results.
\end{acknowledgments}

\appendix

\section{The charged splitting in supersymmetry}
\label{app:susy}

This appendix substantiates the statement of Sec.~\ref{sec:beyondminimal} that a supersymmetric completion at a low scale changes the charged splitting of the doublet, and in a direction that is not unique.

Integrating out a bino and a wino of masses $M_1$ and $M_2$ generates both operators relevant for the splittings. Defining
\begin{equation}
 A\equiv m_Z^2\left(\frac{s_W^2}{M_1}+\frac{c_W^2}{M_2}\right),
\label{eq:Adef}
\end{equation}
one finds for real parameters, $\mu>0$ and at leading order in $\mu/M_{1,2}$,
\begin{align}
 \dmz&\simeq|A| ,
\label{eq:susy}\\
 \dmp^{\rm tree}&\simeq\frac{|A|}{2}+\frac{A\sin2\beta}{2}-\frac{m_W^2\sin2\beta}{M_2} ,
\label{eq:susy2}
\end{align}
which we have checked against a numerical diagonalization of the mass matrices. Two possibilities lead to the sub-MeV neutral splitting required by LZ. If the gaugino masses are of order $m_Z^2/\dmz\sim10^{7}\,\gev$~\cite{FanReece2026,DuWang2026,Yin2026}, the tree-level contribution to the charged splitting is of order $\dmz$ and therefore negligible; the Higgsino then coincides with the minimal doublet. If the gauginos are at the TeV scale, $\dmz$ can be small only through a cancellation between the two terms of Eq.~(\ref{eq:Adef}), which requires $M_1$ and $M_2$ of opposite sign and a tuning at the level of $10^{-3}$--$10^{-4}$~\cite{Cheung2026}. In a scan of $5\times10^5$ points of the phenomenological MSSM, a single point of this kind was found, with a charged splitting of $1\,\gev$~\cite{CMSpMSSM,Bein2026}; supersymmetric interpretations of the event are explored in Refs.~\cite{Bisal2026,Chatto2026,Frolovsky2026,LianYang2026,RadCorr2026}.

For TeV-scale gauginos the last term of Eq.~(\ref{eq:susy2}) is large, but the expansion in $\mu/M_2$ converges poorly and can even give the wrong sign. We therefore diagonalize the tree-level mass matrices exactly along the tuned direction $\dmz\to0$ for $\mu=1.1\,\tev$, and add the radiative splitting. Table~\ref{tab:susy} shows the result. For $M_2\mu<0$ the charged splitting always increases, and the chargino lifetime decreases from the radiative value towards the prompt regime as the wino becomes lighter; the point found in the scan mentioned above belongs to this branch. For $M_2\mu>0$ the splitting increases at large $\tan\beta$ but decreases at small $\tan\beta$, where the chargino travels several centimeters; for $|M_2|\simeq5$--$9\,\tev$ the splitting even drops below the pion mass, and the proper decay length grows to tens of meters. Depending on its boost, the chargino can then produce either a disappearing track or a track that survives through the detector. A low-scale supersymmetric completion can thus move the doublet into the full range of lifetimes illustrated in Fig.~\ref{fig:ctau}. It also adds further superpartners---the wino- and bino-like states themselves and, if they are light enough, colored ones---which provide additional handles at the LHC~\cite{CMSpMSSM,ATLASpMSSM}.

\begin{table}[t]
\caption{Charged splitting $\dmp$ in GeV and decay length $c\tau$ in mm (separated by a slash) of a Higgsino with $\mu=1.1\,\tev$ and TeV-scale gauginos, with splittings from a tree-level diagonalization along the direction $\dmz\to0$, plus the radiative splitting. The indicative lifetimes are evaluated at the displayed, rounded splittings using the doublet one-pion plus leptonic width of Appendix~\ref{app:model}; additional hadronic modes at larger splittings are not included. For very heavy gauginos one recovers $\dmp=0.36\,\gev$ and $c\tau=6$\,mm.}
\label{tab:susy}
\begin{ruledtabular}
\begin{tabular}{lccc}
$|M_2|$ & $\tan\beta=4$ & $\tan\beta=10$ & $\tan\beta=40$ \\
\hline
\multicolumn{4}{l}{$M_2\mu>0$}\\
$3\,\tev$ & $0.21$ / $40$ & $0.55$ / $1.4$ & $0.73$ / $0.52$ \\
$6\,\tev$ & $0.12$ / $2.7\times10^{4}$ & $0.37$ / $5.5$ & $0.49$ / $2.1$ \\
$10\,\tev$ & $0.15$ / $2.2\times10^{2}$ & $0.32$ / $9$ & $0.41$ / $3.9$ \\
$20\,\tev$ & $0.22$ / $33$ & $0.32$ / $9$ & $0.36$ / $6$ \\
\hline
\multicolumn{4}{l}{$M_2\mu<0$}\\
$3\,\tev$ & $1.35$ / $0.046$ & $1.03$ / $0.14$ & $0.85$ / $0.29$ \\
$6\,\tev$ & $0.90$ / $0.24$ & $0.69$ / $0.64$ & $0.57$ / $1.3$ \\
$10\,\tev$ & $0.68$ / $0.67$ & $0.54$ / $1.5$ & $0.46$ / $2.6$ \\
$20\,\tev$ & $0.52$ / $1.7$ & $0.44$ / $3.1$ & $0.39$ / $4.6$ \\
\end{tabular}
\end{ruledtabular}
\end{table}

\section{Details of the FCC-hh yield model}
\label{app:model}

This appendix specifies the model of Sec.~\ref{sec:coverage} and documents its validation.

\emph{Decay widths.} We use the leading heavy-to-heavy charged-current width, neglecting the neutral splitting compared with the charged splitting in the sensitivity windows~\cite{Bottaro2022}. For $Q=\pm1$, define $C_Q=j(j+1)-Y^2+QY$. The leptonic modes are
\begin{align}
 \Gamma_\ell(\Delta)&=C_Q\frac{G_F^2\Delta^5}{15\pi^3}F(m_\ell/\Delta),
 \label{eq:leptonicwidth}\\
 F(x)&=30\int_x^1dy\,y\sqrt{y^2-x^2}(1-y)^2,
 \qquad 0\leq x<1 ,
 \label{eq:leptonicphase}
\end{align}
with $F(0)=1$ and $\Gamma_\ell=0$ for $\Delta\leq m_\ell$. The total width is the sum over $\ell=e,\mu$ plus $C_Q$ times Eq.~(\ref{eq:pionwidth}), with the pion term set to zero below threshold. This prescription gives a continuous lifetime at $m_\pi$ and retains both lepton thresholds. Figure~\ref{fig:ctau} and the sensitivity tables use these same widths. Below all charged-current decay thresholds the charged state is stable in this approximation and its disappearing-track probability vanishes. Neglecting the neutral splitting is justified away from individual decay thresholds. Near the pion threshold the two neutral final states instead open at charged splittings separated by $\dmz$. The very narrow window obtained at the upper fermionic-quintuplet mass in the input variation below requires both channels to be treated separately and should not be assigned sub-MeV physical precision.

\emph{Kinematics.} A pair of invariant mass $M\ge2m$ is produced in association with a jet of transverse momentum $p_\mathrm{T}$, which we take to be central so that the partonic energy is $\sqrt{\hat s}=p_T+\sqrt{p_\mathrm{T}^2+M^2}$. We impose the physical endpoint $\sqrt{\hat s}\leq100\,\tev$ on the integration. The rate is weighted by the quark--antiquark luminosity at this energy, computed with the CT14 parton distributions~\cite{CT14}, by the partonic pair cross section $\hat\sigma(M)\propto\beta(3-\beta^2)/(2M^2)$ for fermions and $\beta^3/(4M^2)$ for scalars, with $\beta=(1-4m^2/M^2)^{1/2}$, and by the soft-emission spectrum $dp_\mathrm{T}/p_\mathrm{T}$. The jet threshold is identified with the missing-momentum requirement of the analysis, $2$ or $4\,\tev$. Each charged state receives the transverse momentum $p_\mathrm{T}^\chi=[(p_\mathrm{T}/2)^2+\tfrac23(\beta M/2)^2]^{1/2}$, which combines the recoil against the jet with the average transverse component of the motion in the pair rest frame, and enters Eq.~(\ref{eq:Ptrack}).

\emph{Couplings and retained channels.} For each pair of multiplet members we use the squared couplings to the $W$ boson, $g^4[j(j+1)-T_3(T_3+1)]/4$, and to the neutral gauge bosons in the unbroken phase, $(g^2T_3T_3^q+g'^2YY^q)^2$, summed over quark chiralities and weighted by the up- and down-quark content of the proton. The tabulated estimate retains the production channels feeding one singly charged gauge branch and its antiparticle. States with $|Q|>1$ are treated as prompt cascades into the singly charged branch of the same sign. The other singly charged branch is omitted from that estimate, not assumed to decay promptly. Relative to the doublet width, the $Q=-1,+1$ factors are $(3,4)$ for the quadruplet and $(4,6)$ for the quintuplet. In each signal region we evaluate the retained channels for both possible assignments of the lightest branch, take the smaller yield, and then select the region with the stronger expected exclusion using the lower normalization. The lifetime in Table~\ref{tab:coverage} belongs to that retained branch. For two tracks with individual probabilities $P_i$ and $P_j$, the geometrical probability to reconstruct at least one is
\begin{equation}
 P_{ij}=1-(1-P_i)(1-P_j).
\label{eq:twotrack}
\end{equation}

\emph{The second singly charged state.} A heavier singly charged state can also have appreciable track acceptance. For example, in the approximation $\dmz=\eta=0$, both charged splittings approach $167\,\mev$ at the spectral upper bound. The two quadruplet proper decay lengths are then approximately $36$ and $27\,\mathrm{mm}$, and the quintuplet values are $27$ and $18\,\mathrm{mm}$. Neither state should be classified as prompt solely because it is not the lightest.

The reason for retaining the smaller single-branch yield is precise in the limit of negligible mixing. Write $D=2\delta_g+\dmz$ and let $S_\pm(d)$ be the yield from channels feeding the $Q=\pm1$ branch with splitting $d$, at a fixed normalization and in a fixed signal region. The two assignments of the lightest branch give total yields
\begin{align}
 S_{\rm all}^{(-)}(d)&=S_-(d)+S_+(D-d),\nonumber\\
 S_{\rm all}^{(+)}(d)&=S_+(d)+S_-(D-d).
\label{eq:bothbranches}
\end{align}
These channel sets are disjoint for the representations considered here. Since the omitted yields are nonnegative,
\begin{equation}
 \min[S_-(d),S_+(d)]\leq
 \min[S_{\rm all}^{(-)}(d),S_{\rm all}^{(+)}(d)].
\label{eq:branchlower}
\end{equation}
This is a lower estimate within the same production, cascade and selection model; it requires an inclusive one-track selection whose acceptance is not reduced by an additional charged track. It is not a detector-independent lower bound. As a numerical check, at $m=4.8\,\tev$ and $d=150\,\mev$, setting $\dmz=0$ consistently with the yield tables gives the other splitting as $184\,\mev$. In the $\met>4\,\tev$ region with the lower normalization, the two assignments give $30.38$ and $34.37$ events after adding the second branch. The retained estimate, $29.90$ events, is smaller than either total.

\emph{Charged mixing near the endpoint.} The gauge-branch treatment requires $|b|\gg|\eta|$ in Eq.~(\ref{eq:chargedmatrix}); the mixing angle satisfies $|\tan2\theta|=|\eta/b|$. This condition can fail within a splitting interval of order $\dmz$ near the upper bound, even though $\dmz$ is much smaller than each charged-neutral splitting. Both charged mass eigenstates must then be propagated, with production couplings and decay amplitudes rotated into their mass basis and with Eq.~(\ref{eq:twotrack}) applied to each channel and cascade. Such a treatment was included in the muon-collider study of Ref.~\cite{Bottaro2022}; it is not implemented in our FCC-hh rescaling. Consequently, the $167\,\mev$ rows and the quadruplet endpoint are gauge-basis extrapolations, not established exclusions of the mixed spectrum. The interval fractions in Table~\ref{tab:windows} and Fig.~\ref{fig:coverage} describe the stated single-branch estimate. The sub-pion endpoints and the scalar-quadruplet and fermionic-quintuplet upper endpoints lie well away from this narrow mixing region.

\emph{Normalization and statistics.} The yield is normalized to the $3\,\tev$ wino of Ref.~\cite{Saito2019}, for which $287$ signal and $42.6$ background events are expected in the region with $\met>2\,\tev$. The upper edge of our bands is normalized instead to the $1\,\tev$ Higgsino, with $19.0$ signal and $1.6$ background events for $\met>4\,\tev$. Expected exclusions and discovery significances use a common fractional background uncertainty $f=0.30$ in both signal regions~\cite{Cowan2011,KumarMartin}. In particular, with $v=(fB)^2$, the discovery significance is
\begin{align}
 Z_{\rm disc}^2={}&2\left[(S+B)\ln\frac{(S+B)(B+v)}{B^2+(S+B)v}\right.\nonumber\\
 &\left.\hspace{8mm}-\frac{B^2}{v}\ln\left(1+\frac{vS}{B(B+v)}\right)\right].
\label{eq:Zdisc}
\end{align}
The statistics-only expression is not used for any discovery number quoted here.

\emph{Thermal-mass and radiative-splitting variations.} For each candidate we repeat the charged-splitting scan at the lower, central and upper thermal masses quoted in Table~\ref{tab:multiplets}. We combine these independently with $\delta_g=163,167,171\,\mev$. For $n=4,5$, each scan is clipped to the varied EFT upper bound, and its interval fraction uses that same bound in the denominator. For $n=2,3$, the radiative benchmark is varied by multiplying the central $354$ or $542\,\mev$ splitting by $\delta_g/(167\,\mev)$, as a sensitivity test based on Eq.~(\ref{eq:spectrum}). This prescription does not replace a higher-order spectrum calculation. All other production, width, detector and statistical inputs are held fixed. Table~\ref{tab:inputvariation} reports the envelope of these nine combinations separately for the two yield normalizations; it is not a confidence band or a model of correlated theory errors.

For the radiative scalar doublet this gives $\mu_{95}=0.24$--$0.48$ with the conservative normalization, or $0.067$--$0.135$ with the optimistic one. For the fermionic doublet the corresponding ranges are $0.78$--$2.08$ and $0.22$--$0.59$. The radiative triplets remain far outside the track reach. At $m=10.6\,\tev$ the conservative quintuplet calculation gives $139.5705<\dmin/\mev<139.7281$, a window of about $0.16\,\mev$ for the input $m_\pi=139.57\,\mev$. Doubling the integration grid changes that width by about $0.2\%$, but the omitted neutral-channel threshold separation and detector effects are more consequential. 

\begin{table*}[t]
\caption{Variation of the FCC-hh track-window endpoints with the thermal masses in Table~\ref{tab:multiplets} and $\delta_g=163,167,171\,\mev$. Each endpoint column gives the spread of that endpoint across the nine input combinations, not a single exclusion interval. C and O denote the conservative and optimistic yield normalizations. The last two columns give the corresponding ranges of interval-length fractions for the larger multiplets, using the varied full EFT interval. These are theory-input variations, not confidence intervals; the gauge-branch and detector qualifications of the central estimate still apply.}
\label{tab:inputvariation}
\footnotesize
\begin{ruledtabular}
\begin{tabular}{lcccccc}
Candidate & \multicolumn{2}{c}{C: endpoint spread [MeV]} & \multicolumn{2}{c}{O: endpoint spread [MeV]} & \multicolumn{2}{c}{Fraction [\%]} \\
& lower & upper & lower & upper & C & O \\
\hline
$2^{S}_{1/2}$ & $87.93$--$88.31$ & $383.22$--$387.59$ & $68.29$--$68.58$ & $416.47$--$421.06$ & --- & --- \\
$2^{F}_{1/2}$ & $63.55$--$64.03$ & $346.80$--$351.02$ & $49.37$--$49.74$ & $373.55$--$377.97$ & --- & --- \\
$3^{S}_{1}$ & $77.87$--$81.66$ & $224.32$--$232.21$ & $60.41$--$63.33$ & $243.97$--$252.12$ & --- & --- \\
$3^{F}_{1}$ & $54.37$--$57.26$ & $225.11$--$232.80$ & $42.23$--$44.47$ & $241.66$--$249.67$ & --- & --- \\
$4^{F}_{1/2}$ & $67.61$--$73.46$ & $163.00$--$171.00$ & $52.45$--$56.94$ & $163.00$--$171.00$ & $53.72$--$60.46$ & $65.06$--$69.32$ \\
$4^{S}_{1/2}$ & $123.62$--$136.21$ & $145.83$--$150.95$ & $93.90$--$102.08$ & $158.10$--$163.38$ & $5.63$--$16.77$ & $32.76$--$42.39$ \\
$5^{F}_{1}$ & $111.10$--$139.57$ & $139.73$--$141.52$ & $83.37$--$100.01$ & $142.35$--$146.40$ & $0.09$--$18.66$ & $24.76$--$38.67$ \\
$5^{S}_{1}$ & --- & --- & --- & --- & $0$ & $0$ \\
\end{tabular}
\end{ruledtabular}
\end{table*}

\emph{Validation.} Table~\ref{tab:validation} compares the model with published results. The production and acceptance ratios agree at the level of $15\%$ or better; the last line instead defines the factor-$3.5$ normalization band. These checks do not validate discovery significances. For example, applying Eq.~(\ref{eq:Zdisc}) to the published benchmark yields gives $Z_{\rm disc}=9.48$ for the $3\,\tev$ wino and $6.88$ for the $1\,\tev$ Higgsino. At $1.1$ and $1.2\,\tev$ the Higgsino-normalized model gives $4.93$ and $3.54$. In the fixed $\met>2\,\tev$ region the wino-normalized model reaches $5\sigma$ at $3.69\,\tev$; this is not a reproduction of the approximately $4.5\,\tev$ published reach, which involves the analysis configurations of Ref.~\cite{Saito2019}. The monojet entry refers to the two published $95\%$~CL reach points~\cite{HanMukhoWang}, which should correspond to equal accepted rates if the model describes the mass and representation dependence correctly.

\begin{table}[t]
\caption{Production-rate and acceptance checks of the yield model against published results. ``Layout dependence'' is the ratio of signal acceptances for the three tracker layouts of Ref.~\cite{Saito2019}, and the last line the ratio of the Higgsino and wino yields in their respective signal regions.}
\label{tab:validation}
\footnotesize
\begin{ruledtabular}
\begin{tabular}{lcc}
Quantity & Model & Published \\
\hline
$\sigma_{\widetilde H}/\sigma_{\widetilde W}$, equal mass~\cite{AHHMW} & $0.52$ & $0.52$ \\
$\sigma(1\,\tev)/\sigma(3\,\tev)$~\cite{AHHMW} & $74$ & $70$ \\
$\sigma(10\,\tev)/\sigma(3\,\tev)$~\cite{AHHMW} & $1.3\times10^{-3}$ & $1.5\times10^{-3}$ \\
Layout dependence, wino~\cite{Saito2019} & $0.23:0.53:1$ & $0.23:0.52:1$ \\
Layout dependence, Higgsino~\cite{Saito2019} & $0.05:0.25:1$ & $0.05:0.26:1$ \\
Monojet anchors, ratio~\cite{HanMukhoWang} & $1.05$--$1.14$ & $1$ \\
Yield ratio of the two regions~\cite{Saito2019} & $0.019$ & $0.066$ \\
\end{tabular}
\end{ruledtabular}
\end{table}

\bibliographystyle{apsrev4-2}
\bibliography{references}

\end{document}